\documentclass[10pt,twocolumn]{wlscirep}
\pdfoutput=1 
\usepackage{amssymb,amsmath}
\usepackage{graphicx}
\usepackage{xcolor}
\usepackage{subfiles}
\usepackage{hyperref}
\usepackage{listings}
\usepackage{siunitx}
\usepackage{gensymb}
\usepackage{textcomp}
\usepackage{xspace}
\usepackage{cleveref}
\usepackage{float}
\usepackage[normalem]{ulem}
\usepackage{caption}
\usepackage{lineno}
\usepackage{etoolbox}
\usepackage{hyperref}
\DeclareMathAlphabet{\mathcal}{OMS}{cmsy}{m}{n}
\DeclareMathAlphabet{\mathbfcal}{OMS}{cmsy}{b}{n}

\definecolor{natureblue}{rgb}{0.282, 0.451, 0.643}
\definecolor{naturebluedark}{rgb}{0.236, 0.329, 0.533}

\hypersetup{
    unicode=false,        
    pdftoolbar=true,        % show Acrobat’s toolbar?
    pdfmenubar=true,        % show Acrobat’s menu?
    pdffitwindow=false,     % window fit to page when opened
    pdfstartview={FitH},    % fits the width of the page to the window
    pdfauthor={none},     % author
    colorlinks=true,       % false: boxed links; true: colored links
    linkcolor=naturebluedark,          % color of internal links
    citecolor=naturebluedark,        % color of links to bibliography
}

\newcommand{\ket}[1] {| #1 \rangle}

\newcommand{\ev}[1] {\langle #1 \rangle}
\newcommand{\logical}[1]{\ket{#1}_L}

\newcommand{\new}[1]{{\color{black} #1}}

\newcommand{\Yb}{\ensuremath{^{171}{\rm Yb}^+}\xspace}
\newcommand{\Ttwostar}{\ensuremath{T_2^*}\xspace}

\title{Fault-Tolerant Operation of a Quantum Error-Correction Code}
\author[1,$\dagger$]{Laird Egan}
\author[2]{Dripto M. Debroy}
\author[1]{Crystal Noel}
\author[1]{Andrew Risinger}
\author[1]{Daiwei Zhu}
\author[1]{Debopriyo Biswas}
\author[3,*]{Michael Newman}
\author[5]{Muyuan Li}
\author[2,3,4,5]{Kenneth R. Brown}
\author[1,2]{Marko Cetina}
\author[1]{Christopher Monroe}

\affil[1]{Joint Quantum Institute, Center for Quantum Information and Computer Science, and Departments of Physics and Electrical and Computer Engineering, University of Maryland, College Park, MD 20742}
\affil[2]{Department of Physics, Duke University, Durham, NC 27708}
\affil[3]{Department of Electrical and Computer Engineering, Duke University, Durham, NC 27708}
\affil[4]{Department of Chemistry, Duke University, Durham, NC 27708}
\affil[5]{Schools of Chemistry and Biochemistry and Computational Science and Engineering, Georgia Institute of Technology, Atlanta, GA, 30332}
\affil[*]{Present Address: Google Research, Venice, CA 90291}
\affil[$\dagger$]{Email: laird.egan@gmail.com}

\begin{abstract}
\textbf{Quantum error correction protects fragile quantum information by encoding it into a larger quantum system.  These extra degrees of freedom enable the detection and correction of errors, but also increase the operational complexity of the encoded logical qubit. Fault-tolerant circuits contain the spread of errors while operating the logical qubit, and are essential for realizing error suppression \new{in practice. While fault-tolerant design works in principle, it has not previously been demonstrated in an error-corrected physical system with native noise characteristics.} In this work, we experimentally demonstrate fault-tolerant preparation, measurement, rotation, and stabilizer measurement \new{of a Bacon-Shor logical qubit using 13 trapped ion qubits. When we compare these fault-tolerant protocols to non-fault tolerant protocols, we see significant reductions in the error rates of the logical primitives in the presence of noise. The result of fault-tolerant design is} an average state preparation and measurement error of 0.6\% and a Clifford gate error of 0.3\% after error correction. \new{Additionally, we prepare magic states with fidelities exceeding the distillation threshold, demonstrating all of the key single-qubit ingredients required for universal fault-tolerant operation.   These results demonstrate that fault-tolerant circuits enable highly accurate logical primitives in current quantum systems.  With improved two-qubit gates and the use of intermediate measurements, a stabilized logical qubit can be achieved.}} 
\end{abstract}

\begin{document}
\flushbottom
\maketitle

Quantum computers promise to solve models of important physical processes, optimize complex cost functions, and challenge cryptography in ways that are intractable using current computers~\cite{feynman1986quantum, abrams1997simulation, aspuru2005simulated, reiher2017elucidating, shor1999polynomial}. However, realistic quantum component failure rates are typically too high to achieve these goals~\cite{von2020quantum, gidney2019factor}. These applications will therefore likely require quantum error correction schemes to significantly suppress errors~\cite{aharonov2008fault, knill1996threshold}.

Quantum error correcting codes combine multiple physical qubits into \textit{logical} qubits that robustly store information within an entangled state~\cite{gottesman1997stabilizer, shor1995scheme, knill1997theory}. However, these codes are not enough on their own. Fault-tolerant (FT) operations, which limit the ways in which errors can spread throughout the system, must also be used.  Without them, the logical error rate \new{may be limited by faults at critical circuit locations that cascade into logical failures, negating the advantage of error-correction}. 

FT state preparation, detection, and operations have been demonstrated using quantum error detecting codes with four data qubits~\cite{corcoles2015demonstration, takita2017experimental, linke2017fault, harper2019fault, andersen2020repeated}. These codes can identify when errors have occurred, but do not extract enough information to correct them. There have also been quantum demonstrations of classical repetition codes to correct quantum errors restricted along one axis ~\cite{cory1998experimental, chiaverini2004realization, schindler2011experimental, reed2012realization, riste2015detecting, kelly2015state}. In other work, qubits have been encoded into quantum error correcting codes that can correct all single qubit errors, but the encoding procedure was not fault-tolerant \cite{gong2019experimental} and the system was not large enough to measure the error syndromes \new{non-destructively using ancilla}~\cite{nigg2014quantum,luo2020quantum}. Parallel work on bosonic codes has demonstrated encoded operations \cite{heeres2017implementing,fluhmann2019encoding}, fault-tolerant detection, one-axis~\cite{ofek2016extending}, and two-axis~\cite{campagne2020quantum} error correction on encoded qubits. For both qubit codes and bosonic codes, fault-tolerant state preparation of a code capable of correcting all single-qubit errors has not been achieved.

\begin{figure*}[ht!]
\centering
\includegraphics[width=\textwidth]{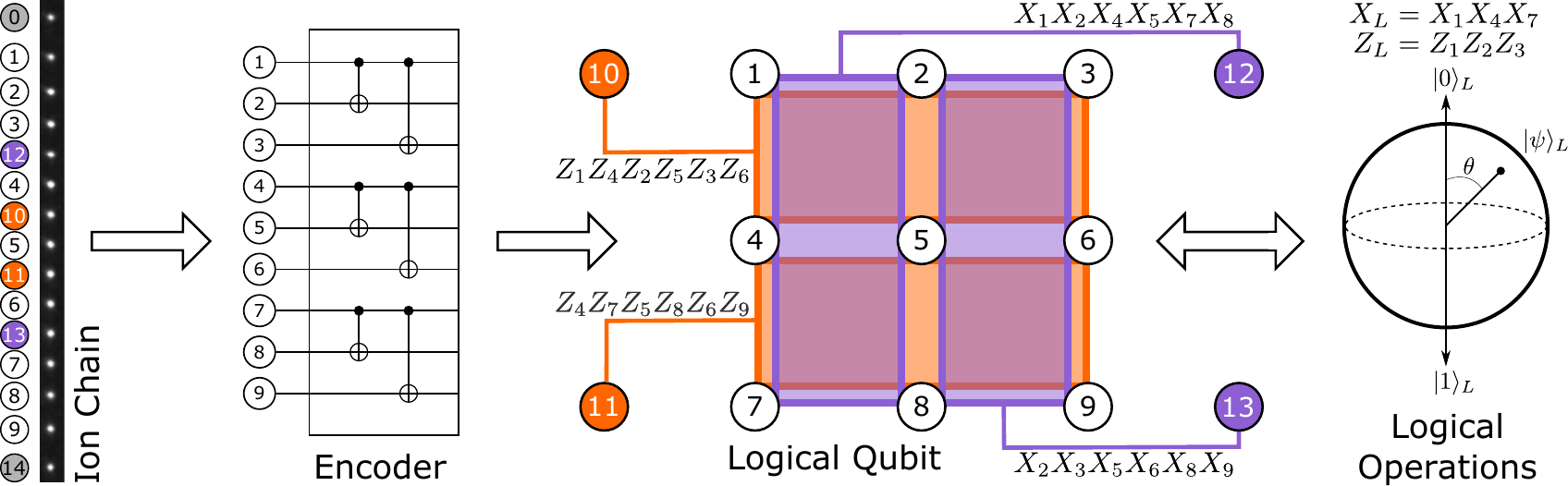}
    \caption{\textbf{The Bacon-Shor subsystem code implemented on a 15 ion chain.} Bacon-Shor is a [[9,1,3]] subsystem code that encodes 9 data qubits into 1 logical qubit. Four weight-6 stabilizers are mapped to ancillary qubits $10$, $11$, $12$, and $13$, for measuring errors in the $X$ and $Z$ basis. We demonstrate encoding of the logical qubit, with subsequent \new{logical gate operations or error syndrome extraction.}}
\label{fig1}
\end{figure*}

\new{Here, we present the fully fault-tolerant operation of a quantum error-correcting code. We demonstrate all of the primitives required for FT operation of the encoded qubit: FT preparation, FT measurement, FT logical gates, and FT stabilizer measurement.  Unlike previous works, this code protects against any single circuit fault (along any axis and without postselection), realizing quadratic error suppression \emph{in principle}.  In practice, this error suppression requires high-fidelity components and localized errors to take effect. Remarkably, we observe a logical operation, preparation and measurement in the $Z$-basis that exceeds the performance of its physical counterpart. More generally, we realize high-accuracy logical primitives that outperform the limiting physical operation used in their construction, namely our native two-qubit entangling gate. 

\new{To experimentally verify the properties of fault-tolerance, we compare non-fault-tolerant (nFT) preparation, nFT logical gates, and nFT stabilizer measurement to their FT counterparts and observe the reduced suppression of errors. In the process, we generate high-fidelity encoded magic states above the distillation threshold, which are a critical resource for certain universal FT quantum computing architectures~\cite{bravyi2005universal}.}

We achieve these results by matching a versatile quantum error-correcting code to the unique capabilities of a state-of-the-art ion trap quantum computer. The ion trap system is simultaneously large enough to run fault-tolerant primitives -- which require more qubits than the error-correction codes themselves -- while remaining accurate enough to realize high-fidelity encoded operations. We leverage the all-to-all connectivity of the device to implement a subsystem quantum error-correction code that does not require intermediate measurement to achieve fault-tolerance.  This allows us to study simple primitives in error-correction codes, even without repeated stabilization required for long-lived memories.}

The quantum computer used in this work consists of laser-cooled \Yb ions trapped above a microfabricated chip~\cite{Maunz2016} in a room-temperature vacuum chamber. Each physical qubit is encoded in the $^2$S$_{1/2}$ electronic ground state hyperfine ``clock" states of a single \Yb ion, $\ket{0} \equiv \ket{F=0;m_F=0}$, $\ket{1} \equiv \ket{F=1;m_F=0}$, with a qubit frequency splitting of $\omega_0 = 2\pi \times 12.642820424(4)$~ GHz. The qubits have a measured $T_2$ decoherence time in excess of $2.75$~s (limited by the stability of external magnetic fields) and average single-shot detection fidelity of $>99.5\%$. Quantum gates are driven by individually optically addressing up to 32 equispaced ions in a single chain via a multi-channel acousto-optic modulator (AOM)~\cite{DebnathQC:2016}. We implement high-fidelity native single-qubit and two-qubit gates with fidelities of 99.98\% and 98.5-99.3\%, respectively. All-to-all two-qubit gate connectivity is achieved through coupling of ions via a shared motional bus~\cite{wright2019benchmarking}. \new{Details of the system, characterization, and benchmarking are available in the Methods and Supplementary Information.}

As shown in \crefformat{figure}{Fig.~#2#1{}#3}\cref{fig1}, we implement a [[9,1,3]] Bacon-Shor code~\cite{BaconBaconShor2006, aliferis2007subsystem}. \new{Because it has distance $3$, the code is able to correct any single-qubit error.} This code is well-suited to near-term ion-trap quantum computing architectures \new{for two reasons. First, Bacon-Shor codes can be prepared fault-tolerantly without intermediate measurement. Compared with the typical projective preparation of topological codes, unitary preparation requires fewer gates and less ancillary qubits. This allows us to demonstrate FT primitives with fewer resources and without intermediate measurements. Second, this code choice is a reasonable midpoint between the qubit efficiency of the 7-qubit Steane code and the robustness of the Surface-17 code \cite{debroy2020logical}. Although the Bacon-Shor stabilizers are weight-6 and non-local, they can be fault-tolerantly measured using only one ancilla per stabilizer~\cite{li2018direct} and leverage the all-to-all connectivity in the device.}

\new{As a subsystem code, the Bacon-Shor code is a generalization of Shor's code that has 4 additional degrees of freedom known as gauge qubits~\cite{shor1995scheme}.  For particular choices of gauge, its logical states are products of GHZ states:}
\begin{equation}
    \begin{aligned}
    \logical{0/1} \otimes \ket{\overline{X}}_G &= \frac{1}{2\sqrt{2}} (|+++\rangle \pm |---\rangle)^{\otimes 3},\\
    \logical{+/-}\otimes \ket{\overline{Z}}_G  &= \frac{1}{2\sqrt{2}} (|000\rangle \pm |111\rangle)^{\otimes 3},\\
    \end{aligned}
    \label{eq: BS logical states}
\end{equation}
where $\ket{\pm}=(\ket{0}\pm\ket{1})/\sqrt{2}$ and $\ket{\overline{X}/\overline{Z}}_G $ refer to different states of the gauge qubits (see Supplementary Information).

Bacon-Shor codes support a wide range of FT operations, \new{including state preparation, state measurement, gates, and stabilizer measurement}. Fault-tolerance, as a design principle, ensures faults on physical operations do not propagate to uncorrectable multi-qubit failures in the circuit. \new{As seen in Eq.~\ref{eq: BS logical states}, not all Bacon-Shor logical states require global entanglement. It is precisely this decomposition into decoupled GHZ states that allows Bacon-Shor to be prepared unitarily and fault-tolerantly. In the $Z_L$/$X_L$ basis, the logical information is encoded redundantly into the relative phase of each state. While a single circuit fault may corrupt one of the three GHZ states, the information can be recovered from the other two. 

FT measurement (in the $X$/$Z$ basis) is performed by individually measuring the data qubits (in the $X$/$Z$ basis). From this information, one can recover relevant stabilizer outcomes as correlations among the single data-qubit outcomes.  This post-processed information is then combined with any previously extracted syndromes, and then collectively decoded to produce a correction. It is worth emphasizing that, although our system does not currently support intermediate measurements, this post-processing step does not differ from the final step of a logical qubit memory experiment with multiple rounds of intermediate measurements. 

Fault-tolerance in logical gates is often achieved via \textit{transversal gates}, which are physical operations that act independently on each qubit in a code block. Bacon-Shor codes have transversal constructions, when allowing permutations, for $\{CNOT_L,\:H_L,\: Y_L\left(\pi/2\right),\: X_L\}$~\cite{shor1996fault,terhal2015quantum, dennis2002topological}. Here, $Y(\theta)$ indicates exponentiation of the Pauli-$\bar{Y}$ matrix, $e^{-i \theta \bar{Y}/2}$. FT non-Clifford logical gates, which are required for universality, can be achieved through magic state distillation~\cite{bravyi2005universal}. 

Finally, measuring error syndromes requires interacting ancillae with multiple data qubits, which could cause damaging correlated errors. However, fault-tolerance is achieved by carefully ordering the interactions, so that correlated errors can be reduced to low-weight errors up to a benign transformation of the gauge subsystem~\cite{li2018direct, li20192d}.}

\begin{figure*}[ht!]
\centering
\includegraphics[width=\textwidth]{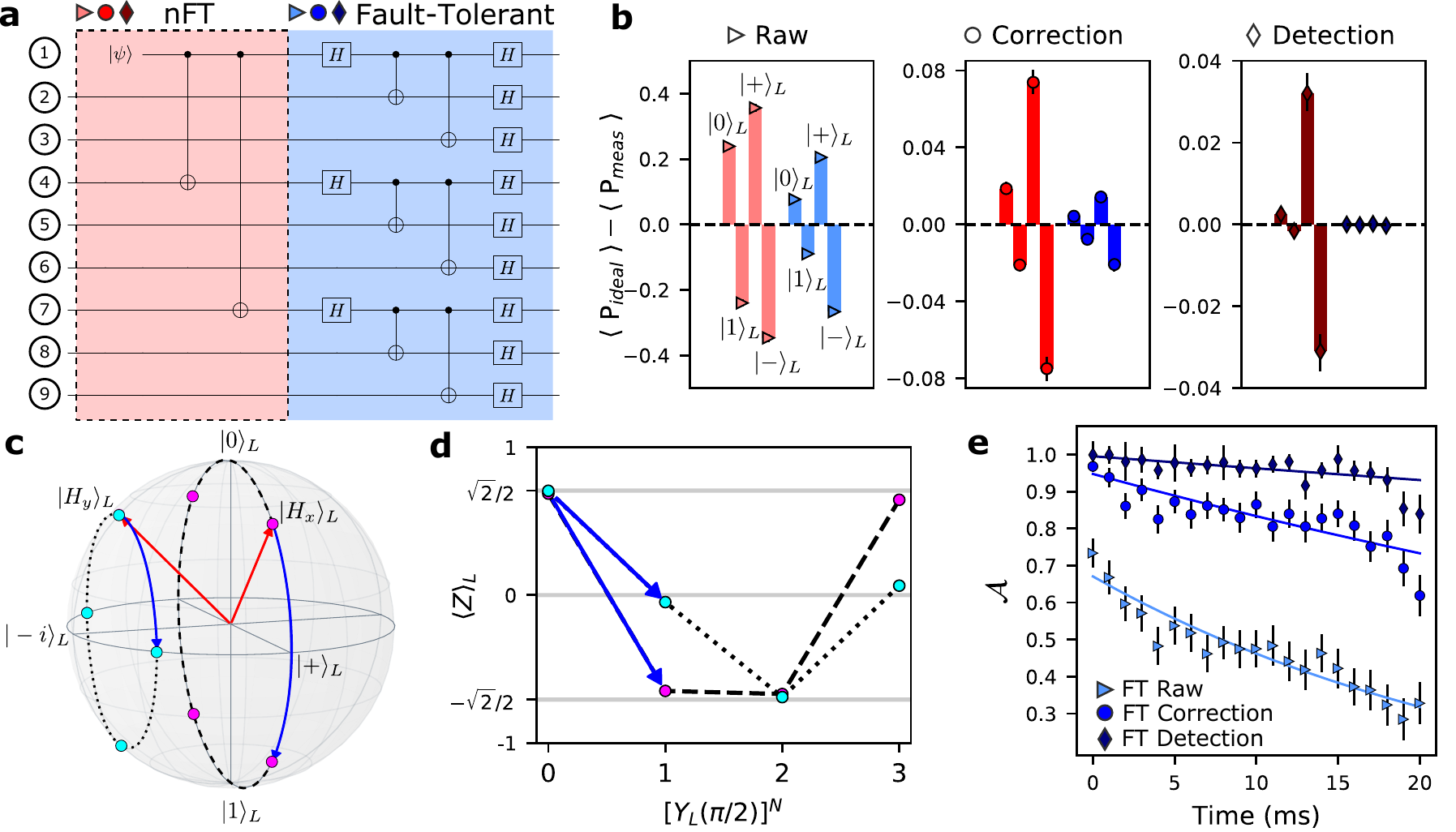}
\caption{\textbf{Fault-tolerant logical qubit state preparation. a,} Encoding circuit for creating logical qubit states. The right subcircuit (blue) is used for FT preparation of $Z$-logical basis states. $X$-logical basis states can be created by omitting the final Hadamard gates. The left subcircuit (red, dashed) can be optionally prepended for nFT preparation of arbitrary logical states. \textbf{b,} Errors for the key basis states of the encoded logical qubit. The measured expectation value of the parity ($P_{meas} = \ev{Z/X}_L$) is compared against the ideal parity of the logical state ($P_{ideal} = \pm 1$). Error bars indicate the $95\%$ binomial proportion confidence interval. \new{\textbf{c,} Magic states $\logical{H_x}$ and $\logical{H_y}$ are directly encoded using the full nFT circuit from \textbf{a} (red arrows). Subsequent $Y_L(\pi/2)$ rotations (blue arrows) are used to bound the fidelity. \textbf{d,} Experimental $\ev{Z}_L$ values for the states depicted in \textbf{c}. $95\%$ binomial proportion confidence intervals are smaller than the data points} \textbf{e,} Logical qubit coherence measured for the $\logical{+}$ state. After each wait time, a varying $Z(\phi)$ gate is applied to every data qubit, followed by $Y_L(-\pi/2)$. A fit of $\ev{X}_L$ depending on $\phi$ to a Ramsey fringe yields the Ramsey amplitude ($\mathcal{A}$). Error bars are the $95\%$ confidence intervals from maximum likelihood estimation fits.The resulting contrast as a function of time is fit to a decaying exponential $Ae^{-t/\Ttwostar}$ for the raw, corrected, and detected data.}
\label{fig2}
\end{figure*}

\section*{Encoding the Logical Qubit}
We embed the 9 data qubits and 4 ancilla qubits of the Bacon-Shor-13 code in a single chain of 15 ions (\crefformat{figure}{Fig.~#2#1{}#3}\cref{fig1}), with the two end ions left idle to obtain uniform spacing of the central 13 ions. The mapping of the code onto the chain is chosen to minimize two-qubit gate crosstalk (details in Supplementary Information). \new{At the start of each experiment, the qubits are initialized to the $\ket{0}$ state. A given circuit is executed by sending appropriate signals to the AOM that implement} single and two-qubit gates on the ion chain with all-to-all connectivity \cite{DebnathQC:2016,wright2019benchmarking}. At the end of a circuit, we perform global state readout by simultaneously collecting state-dependent fluorescence from each ion using high-resolution optics and 32 individual photo-multiplier tubes.

The encoding circuit used to create logical states is shown in \crefformat{figure}{Fig.~#2#1{(a)}#3}\cref{fig2}. The right sub-circuit (blue) is FT because there are no entangling operations between independent GHZ states that would allow errors to propagate; however it is limited to preparation of only $Z$ and $X$ basis states. One may prepend an optional sub-circuit (red, dashed) that enables the encoding of arbitrary \new{$\logical{\psi}$ states, controlled by a single physical qubit state $\ket{\psi}$}. This circuit can produce global entanglement, and allows the possibility of early errors spreading between the separate GHZ states. As a consequence, this circuit loses the FT properties of the $X$ and $Z$ basis preparation circuits. To directly investigate the properties of fault-tolerance, we compare the encoding performance of the right FT sub-circuit to the full nFT circuit with \new{$\ket{\psi} \in \{\ket{0},\ket{1},\ket{+},\ket{-} \}$}.

\new{After measuring the data qubits, the logical measurement outcome is determined by calculating the total parity of all the data qubits in the $Z$-basis, $Z_L= Z_1Z_2...Z_8Z_9$}. From Eq.\ref{eq: BS logical states}, the $\logical{0}$ state has even parity ($\ev{Z}_L = 1$) while $\logical{1}$ has odd parity ($\ev{Z}_L = -1$). Similarly, the $\logical{+/-}$ states have even/odd parity \new{in the $X_L$ basis; a $Y_L(-\pi/2)$} operation following the encoding circuit, maps $\ev{X}_L \rightarrow \ev{Z}_L$. The measured \textit{raw} parity compared to the ideal parity of each logical $Z,X$ basis state is presented in \crefformat{figure}{Fig.~#2#1{(b)}#3}\cref{fig2}. \new{In addition to the total raw parity, $Z_L$, the data qubit measurements also provide the eigenvalues of the two stabilizers in the measurement basis. With this information, error \textit{correction} can be applied, which yields an expected quadratic suppression of uncorrelated errors (i.e. corrects any single error). Alternatively, error \textit{detection} is performed by post-selecting experimental shots conditioned on the $+1$-eigenvalues of the stabilizers. This will yield an expected cubic suppression of uncorrelated errors (i.e. detects any pair of errors). Further details of these protocols are given in the Methods.} 

As shown in \crefformat{figure}{Fig.~#2#1{(b)}#3}\cref{fig2}, using the FT circuit (blue) and performing error correction, we prepare $\logical{0},\logical{1},\logical{+},$ and $\logical{-}$ states with respective errors $0.21(4)\%, 0.39(5)\%, 0.71(7)\%,$ and $1.04(9)\%$. \new{We note that the average state preparation and measurement error for a single physical qubit in the Z basis is 0.46(2)\% (Supplementary Information) compared to 0.30(3)\% in the logical qubit}. This is one context in which the logical qubit clearly outperforms our physical qubit. For the nFT circuit (red) the respective errors are $0.93(8)\%, 1.05(9)\%, 3.7(2)\%,$ and $3.8(2)\%$. \new{The error-detection experiment presents particularly strong evidence for fault-tolerance. We observe a remarkable gap in the failures between the nFT and FT protocols: averaged over the basis states, we see 2 failures of FT error-detection over 13,288 post-selected shots, compared with 197 failures over 12,105 post-selected shots when using nFT error-detection. This agrees with a local error model where we expect cubic suppression of FT error-detection, in stark contrast with nFT error-detection, which can fail due to a single circuit fault. The observed two orders-of-magnitude difference lends further evidence that these circuits, which are fault-tolerant in principle, are also fault-tolerant in practice.}

\new{The nFT preparation circuit can also be used to create $\logical{H_x} = e^{-i\pi \bar{Y}/8}\logical{0}$ and $\logical{H_y} = e^{-i\pi \bar{X}/8}\logical{0}$ magic states, which can be distilled to implement FT non-Clifford gates~\cite{bravyi2005universal, reichardt2005quantum}. \crefformat{figure}{Fig.~#2#1{(c)}#3}\cref{fig2} depicts these states on the logical Bloch sphere, and the results are shown in \crefformat{figure}{Fig.~#2#1{(d)}#3}\cref{fig2}. After error correction, the calculated $\logical{H_x}$ encoding fidelity is $97(1)\%$ (analysis in the Supplementary Information), which is above the distillation threshold of $92.4\%$ \cite{reichardt2005quantum}.}

\new{Performance of the logical qubit as a quantum memory can be characterized by measuring the coherence of $\logical{+}$ versus time.} The results of this logical \Ttwostar experiment are presented in \crefformat{figure}{Fig.~#2#1{(e)}#3}\cref{fig2}. For the raw, error correction, and error detection decoding schemes, we measure a \Ttwostar of 27(2)~ms, 78(9)~ms, and 300(90)~ms. The measured \Ttwostar of each independent GHZ state in the logical qubit is almost entirely explained by the measured $\Ttwostar=0.6(1)$~s of the individual physical qubits (see Supplementary Information). \new{Future work that utilizes the gauge degrees of freedom to create decoherence-free subspaces \cite{lidar1998decoherence,kielpinski2001decoherence} within the $\logical{+}$ GHZ states should readily extend the logical \Ttwostar to the physical \Ttwostar. Ultimately, repeated stabilization of the logical qubit over intermediate time scales will be required to achieve a robust quantum memory.} 

\begin{figure}[ht!]
\centering
\includegraphics[width=0.48\textwidth]{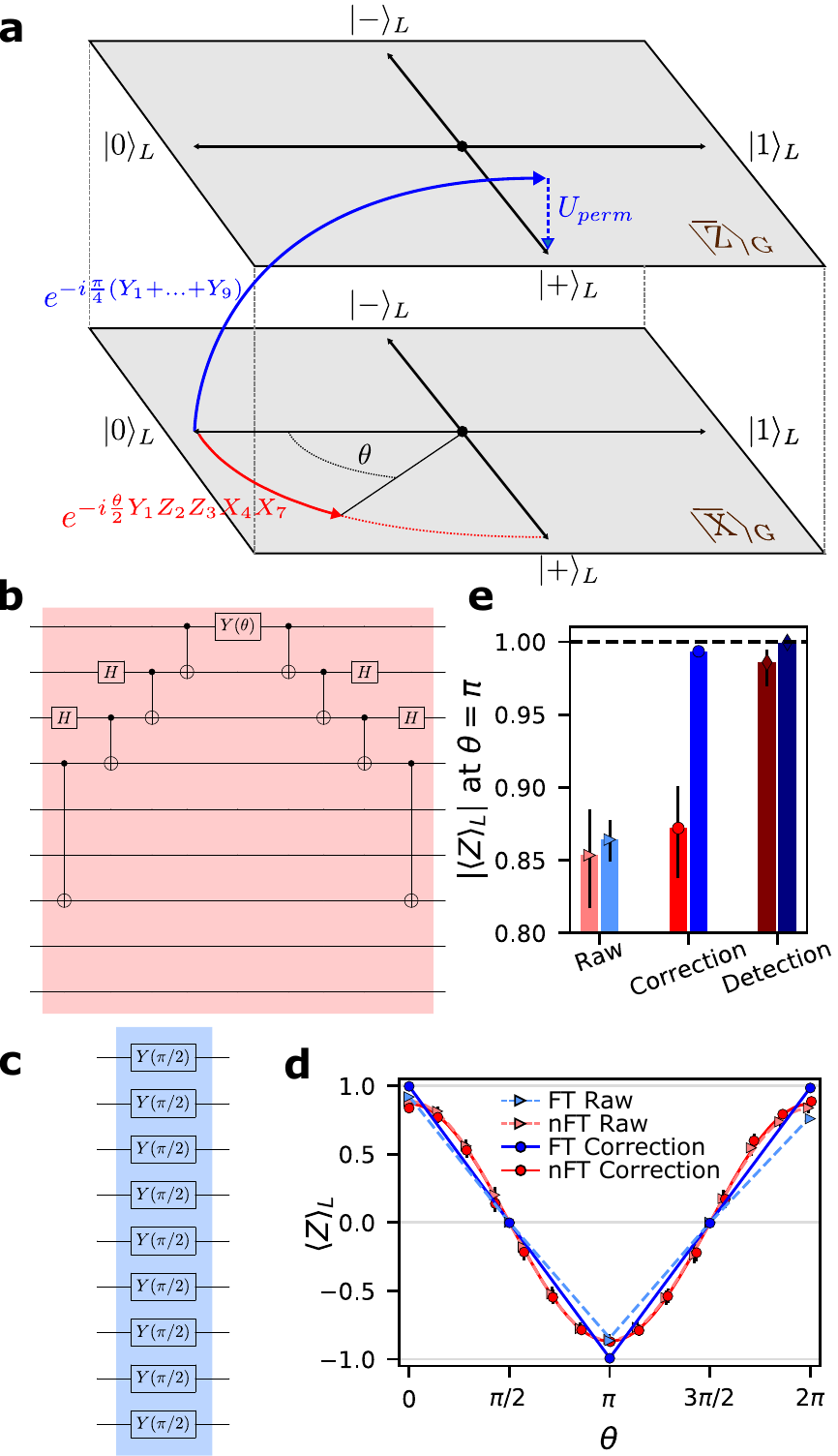}
\caption{\textbf{Manipulating logical states.} \textbf{a,} A schematic depicting different logical operations. A FT discrete logical rotation (blue) operating on $\logical{0}$ is a transversal operation, $Y_L(\pi/2) = Y(\pi/2)^{\otimes9}$, that leaves the code subspace (gray planes) and returns via a permutation of qubit labeling ($U_{perm}$). A nFT continuous logical rotation (red) operating on $\logical{0}$ is a 5-qubit entangling operation, $Y_L(\theta) = Y_1Z_2Z_3X_4X_7(\theta)$, that rotates through the code subspace. At $\theta=\pi/2$, these gates are equivalent up to a gauge transformation. \textbf{b,} The circuit for the nFT gate capable of creating any state along the red curve. \textbf{c,} The circuit for the FT gate shown by the blue curve. \textbf{d,} Experimental results comparing FT (blue) and nFT (red) logical operations. The expectation value of the logical $Z$ operator is fit to a decaying sinusoid $\ev{Z}_L = A\cos(\theta) e^{-\Gamma\theta/\frac{\pi}{2}}$. \new{ \textbf{e,} Detailed view at $\theta=\pi$.} The error bars in \textbf{d/e} are $95\%$ confidence intervals from the binomial distribution. 
}
\label{fig3}
\end{figure}

\section*{Logical Gates}

We implement a $Y_L(\theta)$ rotation on the encoded qubit, which can only be performed transversally for a discrete set of angles~\cite{eastin2009restrictions}. In the case of Bacon-Shor, the smallest transversal $Y_L(\theta)$ rotation we can create is $Y_L(\pi/2)$, which is generated by applying a physical $Y(\pi/2)$ to each data qubit, followed by relabeling the data qubit indices in post-processing (\crefformat{figure}{blue, Fig.~#2#1{a,c}#3}\cref{fig3}). We compare the performance of this FT rotation with a nFT circuit which implements $Y_L(\theta) = Y_1Z_2Z_3X_4X_7(\theta)$ (\crefformat{figure}{red, Fig.~#2#1{a,b}#3}\cref{fig3}). \new{In a perfect system}, these rotations are equivalent for $\theta=N\pi/2$, $N\in \mathbb{Z}$ on the logical qubit, but differ in their operation on the gauge qubits. The nFT gate (\crefformat{figure}{Fig.~#2#1{b}#3}\cref{fig3}) generates entanglement among the separate GHZ states, and so the failure of a single operation in the circuit can lead to the failure of the logical qubit.

The results of these different gate operations on the logical qubit are shown in \crefformat{figure}{Fig.~#2#1{d-e}#3}\cref{fig3}. The gate error per $\pi/2$ angle, corresponding to fit parameter $\Gamma$, is 0.3(1)\% for the FT gate after error-correction. This error rate explains the additional error present for the $\logical{+/-}$ states in \crefformat{figure}{Fig.~#2#1{b}#3}\cref{fig2}, which require two additional $Y_L(\pi/2)$ gates for state preparation and measurement. The rest of the fit values are tabulated in the Supplementary Information. \new{The error at $\theta=\pi$, the maximum gate angle required with optimized circuit compilation, is shown in \crefformat{figure}{Fig.~#2#1{e}#3}\cref{fig3}. The error for the FT gates and nFT continuous rotations is $0.33(18)\%$ and $6.4(1.6)\%$, respectively, after error correction. Compared to the FT circuit, error correction on the nFT rotation provides minimal gains, indicative of a high proportion of weight-2 errors relative to weight-1 errors. In contrast, $\ev{Z}_L$ recovers quite significantly after error detection, indicating that there are still few weight-3 or higher errors in the system. This is a striking example of the value of fault-tolerance, which minimizes the impact of correlated weight-2 errors on the logical qubit}. 

\begin{figure}[t!]
\centering
\includegraphics[width=0.485\textwidth]{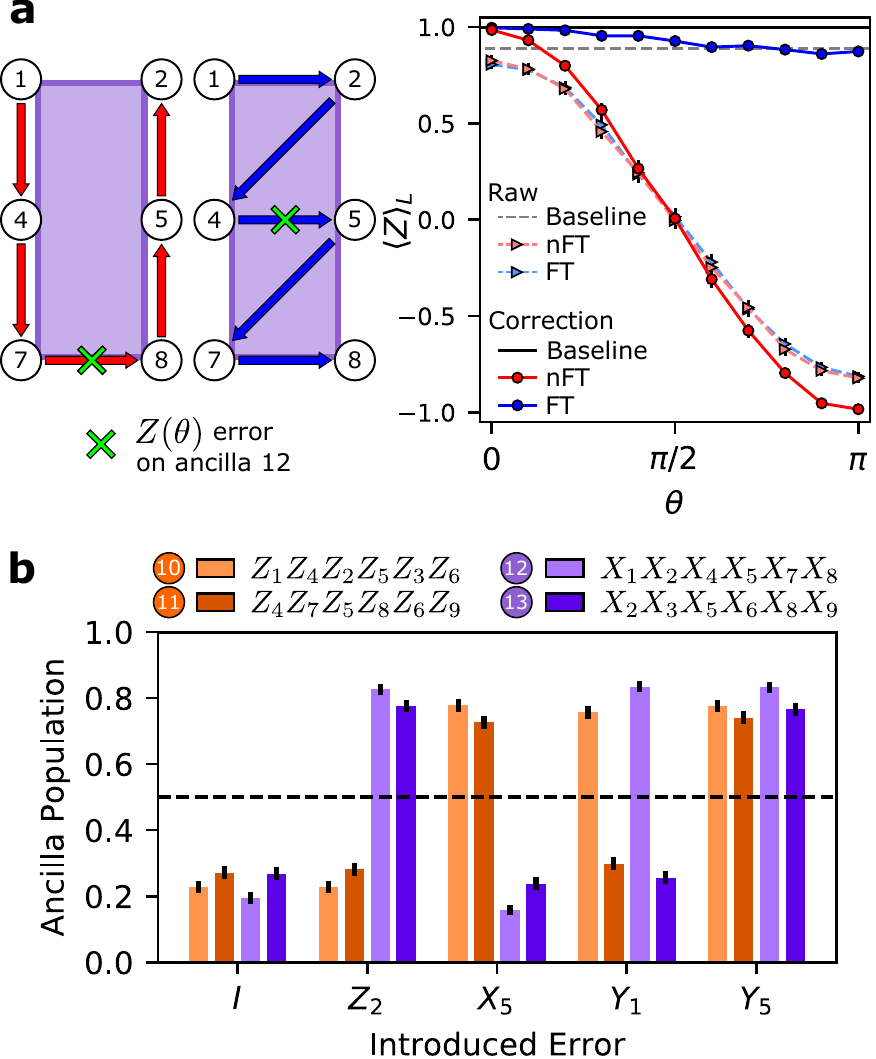}
\caption{\textbf{Detection of arbitrary single-qubit errors.} \new{\textbf{a} Expectation value of the logical $Z$ operator after encoding $\logical{0}$ (Baseline, grey/black line), and then performing the nFT (red) or FT measurement (blue) of a single $X$-type stabilizer with a $Z(\theta)$ error inserted on the ancilla during measurement.} \textbf{b} After encoding $\logical{0}$, different Pauli errors are purposely introduced on a selected data qubit in the code. To detect the error, each stabilizer eigenvalue is mapped onto the state of the corresponding ancilla qubit. The ideal ancilla population is $0/1$ depending on whether an error \textit{did not}$/$\textit{did} anticommute with the stabilizer block. The colored bars correspond to the measured population of the different ancilla qubits. The error bars in \textbf{a,b} indicate the $95\%$ binomial proportion confidence interval. \new{The circuits used to generate the data shown in this figure are given in Extended Data Figure \ref{fig1-ext}.}}
\label{fig4}
\end{figure}

\section*{Stabilizer Measurements}

\new{In stabilizer measurements, fault-tolerance is achieved by a specific ordering of the interactions between the ancilla and the stabilizer block ~\cite{li2018direct}. We insert a variable $Z(\theta)$ error on the ancilla during the measurement of a single stabilizer ($X_1X_2X_4X_5X_7X_8$) and compare the impact of this error in a FT ordering and a nFT ordering. Without correction, a $Z$ error on the ancilla qubit will propagate to an $X$ error on the data qubit and flip $\ev{Z}_L$. 

The results of this experiment are shown in \crefformat{figure}{Fig.~#2#1{a}#3}\cref{fig4}. At the extreme case of $\theta=\pi$, the raw parity is nearly identical in the two cases, but after correction, the FT stabilizer measurement recovers the correct logical parity whereas the nFT stabilizer measurement induces a logical fault. This is because the FT gate ordering propagates a correlated error that decomposes as the product of (at worst) a single qubit fault and a benign transformation of the gauge subsystem. By comparison, the nFT gate ordering propagates a correlated error that directly corrupts the logical subsystem.

At $\theta=0$, (i.e., when no error is added) the error-corrected error rates for $\logical{0}$ after the nFT and FT stabilizer measurement are $0.76(22)\%$ and $0.20(13)\%$, respectively, compared to a baseline encoding error of $0.23(13)\%$. To within statistical error, there is no distinction between performing the FT stabilizer measurement or not, providing strong evidence that this procedure does not corrupt the logical qubit state beyond the error-handling capabilities of the code. On the other hand, there \textit{is} a statistically significant difference ($p$-value $< 0.015$) between the nFT and FT ordering. This again demonstrates the value of fault-tolerance in an apples-to-apples comparison: in two circuits of identical complexity, performing the circuit fault-tolerantly yields an average $4$-times reduction in error. The fact that this reduction is not larger speaks to the precise phase control in our system. While this experiment is specific to $X$ errors propagated from the $X$-stabilizer ancilla qubit, we also characterize $Z$ errors with a similar experiment on the $\logical{+}$ state (Supplementary Information).}

In \crefformat{figure}{Fig.~#2#1{b}#3}\cref{fig4} we show the results of directly measuring the \new{full set of} stabilizers with four additional ancilla qubits. First, the state is fault-tolerantly encoded into the $\logical{0}$ state. Then, an artificial error is applied to a data qubit. \new{Finally, the full set of stabilizers, in sequential order $X$ and then $Z$, are mapped to the ancilla qubits in a single shot}. If no error has occurred, all four stabilizers commute with the logical qubit state and the ancilla qubits should remain in the $\ket{0}$ state. Conversely, if an error did occur on a data qubit, the stabilizers that do not commute with that error flip the state of the ancilla to $\ket{1}$. For example, a Pauli $Y$ error on data qubit 1 anticommutes with both the $X$ and $Z$ stabilizers that measure it, resulting in a flip of ancilla qubits $10$ and $12$, as we observe in the data. This confirms our ability to, \new{ on average, simultaneously} identify arbitrary single qubit errors along both $X$ and $Z$ axes using the stabilizer outcomes.

The data presented in \crefformat{figure}{Fig.~#2#1{b}#3}\cref{fig4} represents a sample of selected errors; the full data set is available in Supplementary Information. Averaged over all the injected errors, the measured ancilla qubits $12$, $13$, $10$, and $11$ (in order of measurement) differ from the expected value by $17.9(3)\%$, $24.8(3)\%$, $24.4(3)\%$, and $29.8(6)\%$, respectively. \new{Most of this non-artificial error is induced by the syndrome extraction circuit itself}. In particular, these results are well explained by the $3.8(2)\%$ raw $\logical{0}$ encoding error, $6.9(5)\%$ error per $X$ stabilizer, $6.4(7)\%$ error per $Z$ stabilizer, and a fixed $7.2(5)\%$ $Z$-type error on the logical qubit that is consistent with the expected raw \Ttwostar-decay over the 3 ms time required to measure $X$ stabilizers, as shown in \crefformat{figure}{Fig.~#2#1{(c)}#3}\cref{fig2}. \new{While these circuits are remarkably accurate given their complexity (30 two-qubit gates in total), we expect that further refinements in gate fidelity are necessary to see improvements over multiple rounds of stabilization.}

\section*{Outlook}
In this work, we have demonstrated \new{high-accuracy fault-tolerant} operation of a logical qubit capable of correcting all single-qubit errors. \new{There are two clear and immediate milestones ahead. One} is to demonstrate a transversal CNOT logical gate that outperforms the physical two-qubit gate, which is the limiting operation in ion systems. This experiment should be possible in the current system given that two-qubit gates on 23 data qubits have recently been demonstrated ~\cite{cetina2020quantum}. \new{The other is to stabilize the state over} multiple rounds of error-correction, which can be achieved by breaking the ion chain to perform mid-circuit detection~\cite{kielpinski2002architecture}. This shuttling will likely require sympathetic cooling schemes, which have been previously demonstrated \cite{home2009complete,pino2020demonstration} and can also be readily implemented in this system~\cite{cetina2020quantum}. 

\section*{Data Availability}
The data that support the findings of this study are available from the corresponding author upon request and with the permission of the US Government sponsors who funded the work.
\vspace{-10pt}
\section*{Code Availability}
The code used for the analyses is available from the corresponding author upon request and with the permission of the US Government sponsors who funded the work.
\vspace{-10pt}
\section*{Acknowledgments}
We acknowledge fruitful discussions with N. M. Linke and the contributions of J. Mizrahi, K. Hudek, J. Amini, K. Beck, and M. Goldman to the experimental setup. This work is supported by the ARO through the IARPA LogiQ program, the NSF STAQ Program, the AFOSR MURIs on Dissipation Engineering in Open Quantum Systems and Quantum Interactive Protocols for Quantum Computation, and the ARO MURI on Modular Quantum Circuits. L. Egan and D. M. Debroy are also funded by NSF award DMR-1747426.
\vspace{-10pt}

\section*{Author Contributions}
L.E. collected and analyzed the data. L.E., D.M.D., C.N., and M.N., wrote the manuscript and designed figures. M.C. and C.M. led construction of the experimental apparatus with contributions from L.E., C.N., A.R., D.Z., and D.B. Theory support was provided by D.M.D., M.N., M.L., and K.R.B.. C.M. and K.R.B. supervised the project. All authors discussed results and contributed to the manuscript.

\section*{Methods}
\subsection*{Experimental implementation}
We trap \Yb in a microfabricated-chip ion trap (High Optical Access 2.1.1 from Sandia National Labs) driven by an RF voltage at a frequency of 36.06 MHz. We define the $x$-axis along the trap axis, with the $z$-axis perpendicular to the chip surface. A magnetic field of 5.183~G along the $z$-axis defines the atomic quantization axis. The individually-addressing (global) Raman beam is oriented along the $z$($y$)-axis of the trap, so that the Raman process transfers momentum to the ions along the $\hat{y}-\hat{z}$ direction. We selectively couple light to the lower-frequency set of radial modes by tilting the trap principal axes using a static electric $yz$ quadrupole. We use quadratic and quartic axial potentials to minimize the spacing inhomogeneity for the middle N-2 ions. In the 15-ion chain, the longest wavelength (in-phase) mode along each trap axis is $(\nu_x,\nu_{y-z},\nu_{y+z}) = (0.193,3.077,3.234)$ MHz. 

An imaging objective with numerical aperture 0.63 (Photon Gear, Inc.) is used to focus each of the 32 individual beams to a waist of $0.85~\mu$m, spaced by $4.43~\mu$m at the ions. The mode-locked 355 nm laser (Coherent Paladin 355-4000) used to drive Raman transitions has been modified to tune the repetition rate of the laser so as to null the 4-photon cross-beam Stark shift. Typical spin-flip Rabi frequencies achieved in our system are 500 kHz. The maximum crosstalk on nearby ions is 2.5\% of the Rabi frequency of the addressed ion.

Before each experiment, the ions are cooled to near the motional ground state through a combination of Doppler cooling and Raman sideband cooling and then initialized into $\ket{0}$ via optical pumping. After the circuit, resonant 369~nm light on the $^2$S$_{1/2} \rightarrow ^2$P$_{1/2}$ cycling transition is used to perform state detection. Scattered light is collected through the 0.63 NA objective and imaged with magnification of 28 onto a multi-mode ($100~\mu$m core) fiber array that is broken out into individual photo-multiplier tubes (Hamamatsu H10682). About 1\% of the total light is detected as counts. Dark/bright states are mapped to $\ket{0}/\ket{1}$ states by setting a threshold at $>1$ photon detected within a detection window (typically $100~\mu$s). State preparation and detection errors are $0.22(2)\%$ and $0.71(4)\%$ for $\ket{0}$ and $\ket{1}$. Detection crosstalk onto neighboring channels is $0.3(2)\%$; see Supplementary Information for detailed error budget.

The entire experiment is controlled by an FPGA (Xilinx) programmed via the ARTIQ software. RF gate waveforms are generated by a 4-channel AWG (Keysight M3202A), one of which drives the global beam, and two of which are routed through a custom switch network onto any of the 15 middle channels of the individual beam AOM at each timestep in the circuit. 

\subsection*{Native ion-trap single-qubit gates}
The native physical single-qubit gate available to our system is a single qubit rotation about a vector in the $xy$-plane, $R(\theta,\phi)$ where $\theta$ is the angle of rotation and $\phi$ is the angle between the rotation axis and the x-axis. In this notation, $RX(\theta) = R(\theta,0)$ and $RY(\theta) = R(\theta,\pi/2)$. Additionally, we use compound SK1 pulses to suppress angle and cross-talk errors~\cite{brown2004arbitrarily}. The SK1 pulses are shaped with a smooth Gaussian amplitude envelope to avoid frequency content that may excite axial motion due to light-induced prompt charge effects from partially exposed semiconductor in the chip trap. Due to hardware limitations, single-qubits gates are run sequentially. We implement virtual $RZ(\theta)$ gates via a software advance of the local oscillator phase, tracked for each individual ion. Before each circuit is run, we calibrate the amplitude of an $RX(\theta)$ on each qubit in the chain. We achieve single-qubit native gate error rates of $1.8(3) \times 10^{-4}$ on a 15-ion chain as measured by randomized benchmarking (see Supplementary Information).

\subsection*{Native ion-trap two-qubit gates}
The native two-qubit operation is the $XX(\theta)$ Ising gate, implemented via a M{\o}lmer-S{\o}rensen interaction~\cite{molmer1999multiparticle}. CNOT gates can be constructed from an $XX(\pi/4)$ gate and additional single qubit gates~\cite{maslov2017basic}. Offline, we calculate laser pulse solutions for $XX$ gates to disentangle the motional modes using amplitude-modulated waveforms \cite{DebnathQC:2016} discretized into 16 segments with linear interpolation between segments to avoid undesirable excitation of the axial motion. In an equispaced chain of 15 ions, we observe that the middle 11 radial motional modes are also roughly equispaced. The laser detuning from motional modes is constant across the waveform and is chosen to sit approximately halfway between two adjacent modes, which leads to particularly simple laser waveforms to eliminate qubit-motion entanglement at the end of the gate. The gate frequency for a particular gate pair is optimized to minimize the required laser power, minimize sensitivity to mode-frequency errors of $<1$~kHz, and to avoid coupling to modes with low spatial frequencies that are subject to heating . Gate durations are 225$~\mu$s. To avoid unwanted couplings, we run two-qubit gates sequentially. Before a batch of circuits is run, we calibrate the amplitude, common phase and differential phase of each gate in the circuit. We achieve between 98.5 and 99.3\% fidelity on a typical gate, measured by parity fringes after a varying odd number of successive non-echoed or echoed $XX$ gates (see Supplementary Information).

\subsection*{Crosstalk Detection}
When available, unused qubits in a circuit are used as flag qubits to detect potential two-qubit gate crosstalk errors. Any experimental shots where an idle qubit is measured in the $\ket{1}$ state are discarded in post-processing. On average, $<4\%$ of the total data is discarded using this method.

\new{
\subsection*{Error Correction Protocol}
Global measurement at the end of each circuit provides the state of all nine data qubits. From this data, we can calculate the raw total parity, $Z_L= Z_1Z_2...Z_8Z_9$, and the eigenvalue of the two Z stabilizers, $S_1=Z_1Z_4Z_2Z_5Z_3Z_6$ and $S_2=Z_4Z_7Z_5Z_8Z_6Z_9$. The processed total parity, $Z_L^\prime$ from the different protocols is then given by the following logic table:}

\begin{table}[H]
\centering
\begin{tabular}{>{\color{black}}c >{\color{black}}c >{\color{black}}c >{\color{black}}c}
    \textbf{Protocol} & \textbf{If} & \textbf{Then} & \textbf{Else}\\
    \hline
    Raw & True & $Z_L^\prime=Z_L$ \\
    Correction & $S_1=-1\parallel S_2=-1$ & $Z_L^\prime=-Z_L$ & $Z_L^\prime=Z_L$\\
    Detection & $S_1=-1\parallel S_2=-1$ & Discard data & $Z_L^\prime=Z_L$\\
\end{tabular}
\end{table}

\onecolumn
\newpage
\renewcommand{\figurename}{Extended Data Figure}

\setcounter{figure}{0}
\crefformat{figure}{Fig.~#2#1{}#3}
\setlength{\intextsep}{8pt}
\setlength{\textfloatsep}{8pt}
\setlength{\floatsep}{8pt}

\begin{figure}[t!]
\centering
\includegraphics[width=\textwidth]{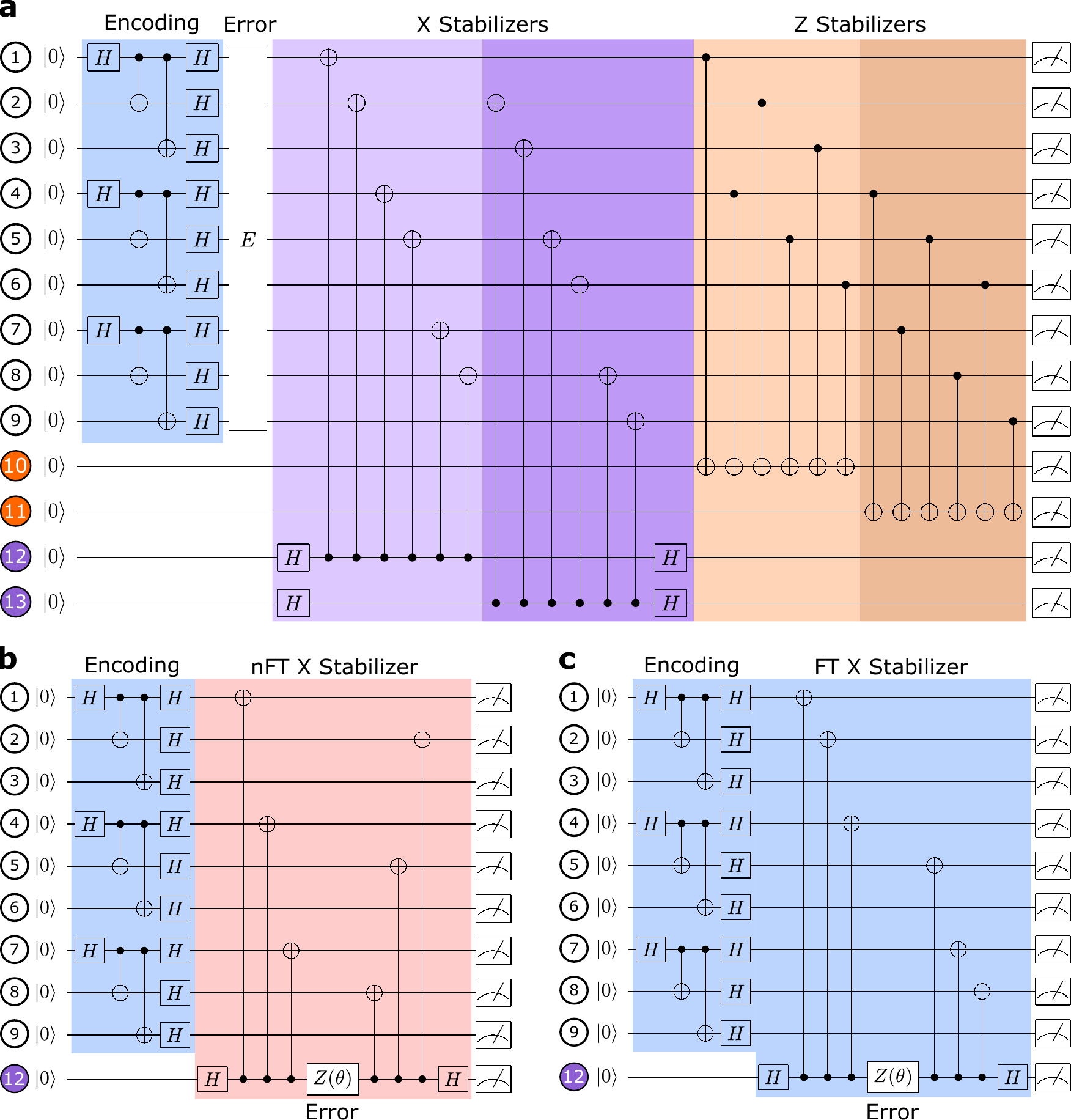}
\caption{\new{\textbf{Stabilizer measurement circuits a,} Direct measurement of the full error syndrome. Various single-qubit "errors" are introduced on any one of the data qubits to generate different ancilla measurement outcomes. This circuit was used to generate the data in Figure \ref{fig4}b of the main text. \textbf{b/c,} Non-fault-tolerant (\textbf{b}, red, right) and fault-tolerant (\textbf{c}, blue, right) stabilizer measurement orderings, performed on a FT-encoded $\logical{0}$ state (\textbf{b/c}, blue, left). In both cases, a variable error $Z(\theta)$ is introduced on the ancilla qubit in the middle of the stabilizer measurement operation. These circuits were used to generate the data in Figure \ref{fig4}a of the main text.}}
\label{fig1-ext}
\end{figure}

\onecolumn
\newpage
\section*{\huge Supplementary Information}

\renewcommand{\thefigure}{S\arabic{figure}}
\renewcommand{\thetable}{S\arabic{table}}
\renewcommand{\theequation}{S\arabic{equation}}
\setcounter{figure}{0}
\crefformat{figure}{Fig.~#2#1{}#3}
\setlength{\intextsep}{8pt}
\setlength{\textfloatsep}{8pt}
\setlength{\floatsep}{8pt}

\subsection*{State preparation and measurement errors}\label{sec:SPAM}
We characterize the state preparation and measurement (SPAM) errors using the following method. We load a single ion and prepare it in the $\ket{0}$ state using optical pumping, from which we may also apply an SK1 \cite{brown2004arbitrarily} $\pi$-rotation to prepare $\ket{1}$. To measure the qubit state, 369~nm light that is resonant with the \{$^2$S$_{1/2}, \rm{F} = 1$\} $\longleftrightarrow$ \{$^2$P$_{1/2}, \rm{F} = 0$\} transition is directed onto the ion and the scattered photons are detected using our array of PMTs. We determine that the ion is bright (dark) when we detect $>1$ ($\leq 1$) photons within a $100~\mu s$ window. We measure a SPAM error of $0.71(4)\%$ when the ion is prepared in $\ket{1}$ (the bright state), and $0.22(2)\%$ for $\ket{0}$ (the dark state), making the average single-qubit SPAM error $0.46(2)$\%. Table~\ref{tab:spam} describes the SPAM error budget, derived either from separate measurements or by fitting Poisson curves to the histogram of photon count event frequency. The measured average detection cross-talk to neighboring PMTs when the target ion is bright is  $0.3(2)\%$.

\begin{table}[h]
\centering
\begin{tabular}{ll}
\textbf{SK1 pulse, 1-state and 0-state error} & \hspace{-.4cm}\textbf{Error budget} \\
SPAM error on bright ion $\equiv \ket{1}$ & \textbf{$0.71\%$} \\
\hspace{.3cm} Bright to dark pumping & $0.55\%$ \\
\hspace{.3cm} Thresholding error & $0.12\%$ \\
\hspace{.3cm} Preparation error (1-qubit randomized benchmarking) \hspace{1.1cm} & $0.03\%$ \\
\vspace{-.25cm} &  \\
SPAM error on dark ion $\equiv \ket{0}$ & \textbf{$0.22\%$} \\
\hspace{.3cm} Dark to bright pumping & $0.13\%$ \\
\hspace{.3cm} Preparation error - incomplete pumping & $0.02\%$ \\
\hspace{.25cm} Background dark counts (measured with no ion qubit) & $0.07\%$ \\
\vspace{-.3cm} & \\
Detection cross-talk error (averaged across neighboring PMTs to bright ion) \hspace{1cm} & \textbf{$0.34\%$}
\end{tabular}
\caption{State preparation and measurement error budget for a single ion in our system. }
\label{tab:spam}
\end{table}

\subsection*{Single qubit gate benchmarking}
\begin{figure}[H]
\centering
\includegraphics[width=0.45\textwidth]{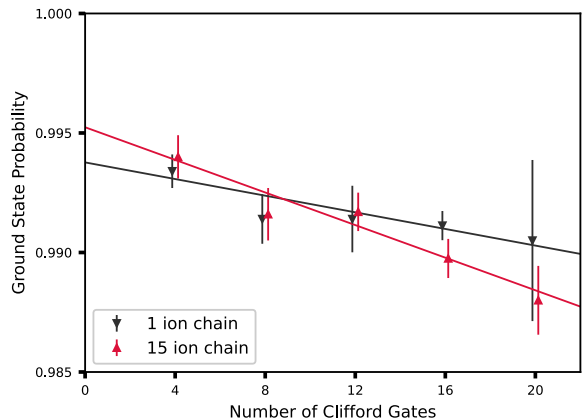}
\caption{\textbf{Randomized benchmarking of single-qubit gates.} The probability to measure a single ion in the ground state after a variable number of Clifford gates in a randomized benchmarking sequence. The slope of the line indicates the per-Clifford fidelity, while the y-intercept indicates the SPAM error. Fit function for 1 ion chain is $0.9938(8)- N * 1.7(7)\times 10^{-4}$, and for 15 ion chain  $0.995(1)- N * 3.4(8)\times10^{-4}$. Error bars shown are the standard error of the mean.}
\label{fig:RB}
\end{figure}

The reported single qubit gate fidelity was measured using single qubit randomized benchmarking \cite{knill2008randomized}, using a sequence of up to 20 random Clifford gates, which were decomposed into our native rotation gates and implemented using SK1 composite pulses. Each random sequence is followed by its inverse in order to, in principle, echo out the gates completely and return the qubit to the initial $\ket{0}$ state. The degree to which the qubit does not return to the initial state quantifies the infidelity of the circuit. The measured occupation of the 
$\ket{0}$ ground state as a function of the number of the applied Clifford gates is shown in \cref{fig:RB}. This benchmarking procedure is performed on a single ion, as well as on an individual qubit in a chain of 15 ions, so as to detect any adverse affects arising from an increase in the system size. The fitted slope of the occupation of the $\ket{0}$ state as a function of the number of the applied Clifford gates indicates a per-Clifford error of $3.4(8) \times 10^{-4}$ on the 15-ion chain, corresponding to an error of $1.8(3) \times 10^{-4}$ per native Pauli gate\cite{Barends2014a}. The offset in the fit is consistent with SPAM errors.

\subsection*{Two-qubit gate performance}
\begin{figure}[H]
\centering
\includegraphics[width=0.55\textwidth]{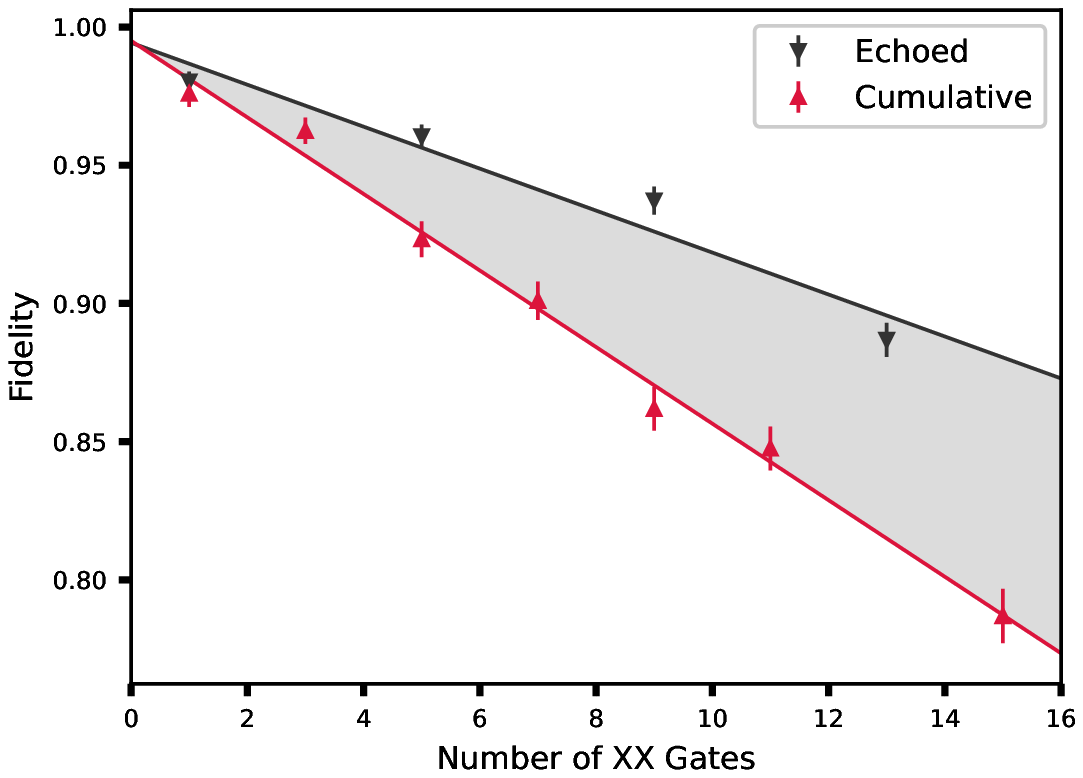}
\caption{\textbf{Two-qubit gate performance.} The cumulative sequence (red) $XX(\pi/4+\epsilon)XX(\pi/4+\epsilon)...$ results in a total error of $N\epsilon$, and the echoed sequence (black) $XX(\pi/4+\epsilon)XX(-\pi/4-\epsilon)...$ cancels this particular type of over/under-rotation error. The true gate fidelity lies somewhere in the shaded region. The slope of the two lines determines the error per gate, or the estimated fidelity to be in the range of $98.5-99.3\%$. Errors are plotted for $95\%$ confidence intervals.}
\label{fig:xxecho}
\end{figure}

Two qubit gate performance was estimated using the results of two gate sequences. On gates with low crosstalk (details in the following section), we anticipate the dominant error in the $XX(\pi/4)$ gate will be an over or under-rotation error by a small angle $\epsilon$ resulting in $XX(\pi/4+\epsilon)$. A sequence of $N$ successive applications of the gate will then will result in an accumulated error of $N\epsilon$. If the phase of the gate is flipped with each successive application, $XX(\pi/4+\epsilon)XX(-\pi/4-\epsilon)...$, then this particular error is suppressed to the extent that it is stable between applications. We also note that the echo sequence will also suppress other forms of coherent errors, such as gate crosstalk. We take the echoed sequence to be the "best-case" scenario and the non-echoed sequence to be the "worst-case" scenario. Within a circuit, we expect the true fidelity of a single $XX$ gate to fall between these two extremes. The fidelity is determined by the parity fringe method\cite{sackett_experimental_2000}. We increase the number of $XX$ gates in the sequence, and take the slope of the fidelity to be the error per gate. The results of this estimation are shown for a single gate between ions 2 and 3 in the chain of 15 in \cref{fig:xxecho}. Thus, the estimated fidelity of a single $XX$ gate is bounded within $98.5-99.3\%$. In other gates, a decrease in fidelity relative to this estimate is primarily due to effects of crosstalk.

\new{
\subsection*{Physical Error Modeling}
\begin{figure}[H]
\centering
\includegraphics[width=0.9\textwidth]{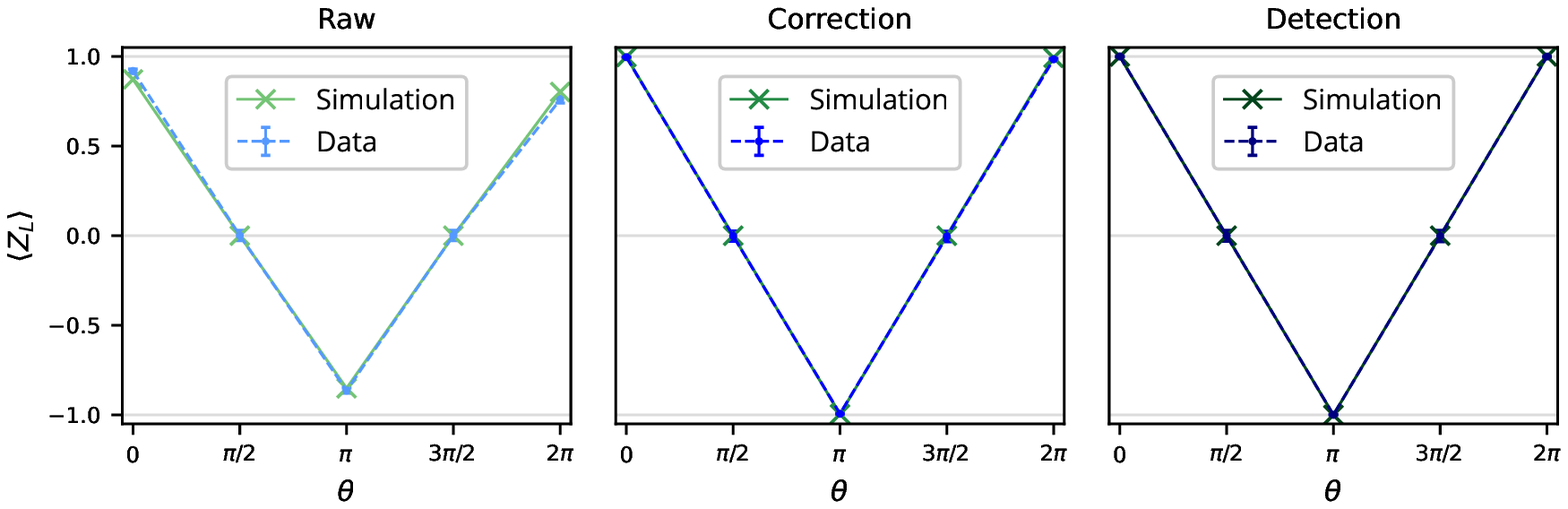}
\caption{\textbf{Matching of simulation and experiment.} The simulated and experimentally measured data for repeated applications of the fault-tolerant logical rotation gate applied to a state initially prepared in $\logical{0}$. The data points labeled "data" are the same as the blue curves shown in Fig 3d of the main text.}
\label{fig:simmatch}
\end{figure}

To confirm our understanding of the experimental system, we find it useful to design physical error models that can replicate our experimental results in simulation. A simple coherent overrotation error model, combined with stochastic measurement and preparation errors closely matches our results, as shown in \cref{fig:simmatch}. The coherent overrotation error channel for a rotation gate $G$ is modeled as
\begin{equation}
    \varepsilon_G(\rho) = \exp (-i\theta G)\rho\exp (i\theta G),
\end{equation}
where $\theta$ is the angle of overrotation. This model is applied to both the M{\o}lmer-S{\o}rensen gate and the single qubit rotation gates, where the fidelities presented in the main text are directly translated into values for $\theta$. These two errors are demonstrated in \cref{fig:errormatch}. The simulation then includes stochastic measurement errors and preparation errors following the SPAM error rates determined from our benchmarking experiments presented in the SPAM error benchmarking section above. Using this four parameter error model, we can capture much of the performance of our system. 

\begin{figure}[H]
\centering
\includegraphics[width=0.9\textwidth]{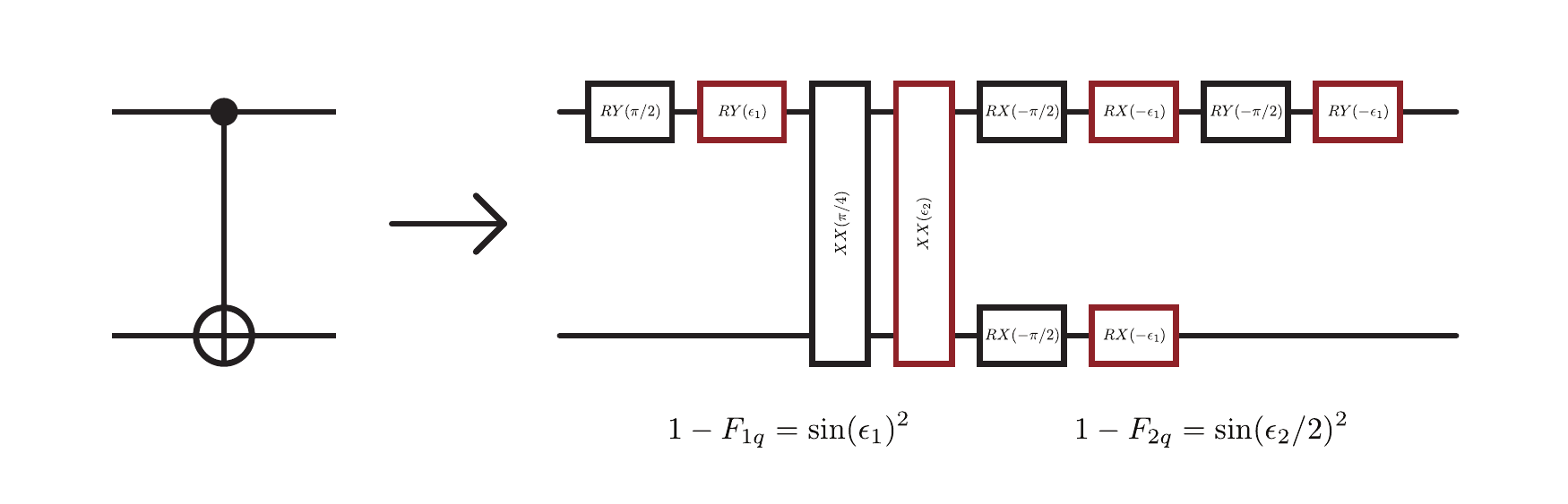}
\caption{\textbf{Gate error model for CNOT.} One possible choice of native gate decomposition (black) and gate errors (red) for a CNOT gate.}
\label{fig:errormatch}
\end{figure}
}

\subsection*{Circuit optimization for crosstalk}
There are several factors to consider when mapping the Bacon-Shor code onto a chain of 15 ions, as shown in Fig. 1 of the main text. In general, ion chains feature all-to-all two-qubit gate connectivity; however, some gates require more optical power than others to achieve maximal entanglement. Considering errors that scale with intensity, such as crosstalk, then gate fidelity is expected to decrease with increasing power requirements. These differences in power requirements can be understood by examining the mode participation symmetries in the chain. For example, ion 8, the center ion, requires high power in nearly all of its gates because it only participates in the even spatial modes (i.e., $b_{8,2n} = 0$, $n=1,2,...,7$ where $b_{i,1}$ is mode-participation factor of the highest-frequency in-phase radial mode for ion $i$). So on average, for a fixed gate frequency, the modes that drive entanglement are further detuned from the gate. We note that this is unique to our choice of amplitude modulated (AM) gates with a fixed frequency; phase/frequency-modulated (PM/FM) gates or multi-tone gates may have different chain symmetry considerations.

\begin{figure}[H]
\centering
\includegraphics[width=0.5\textwidth]{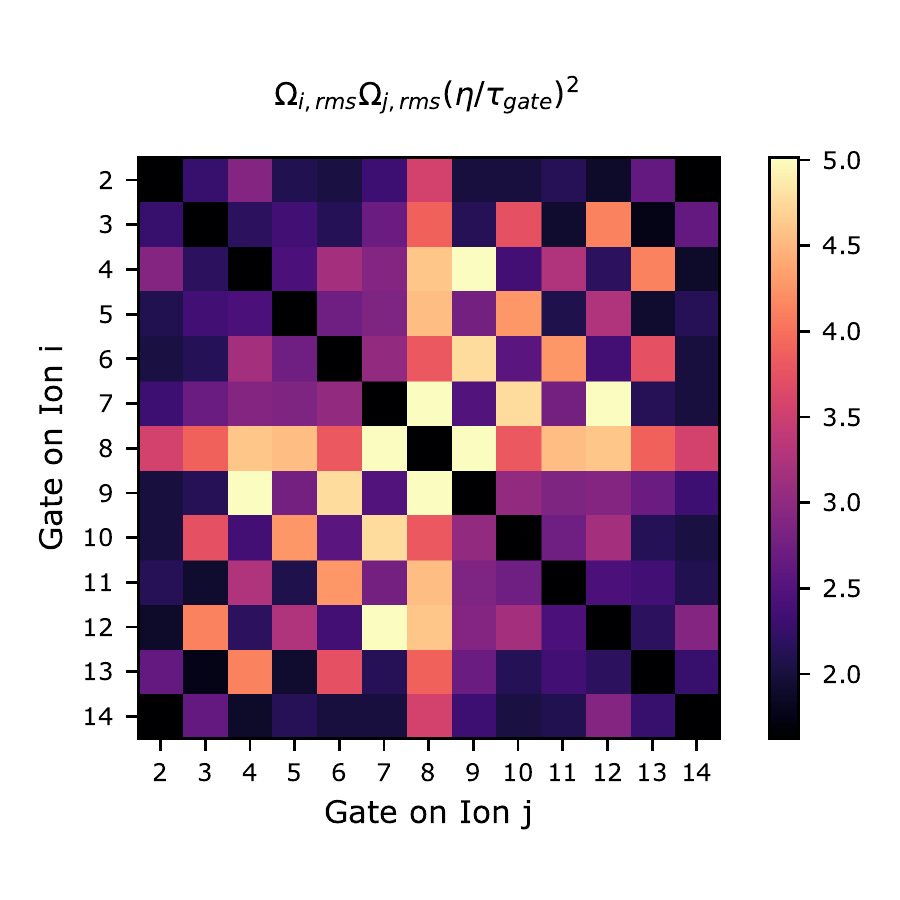}
\caption{\textbf{Power requirements for XX gates in our system.} For an XX gate on ions $i,j$, the (RMS) Rabi frequency ($\Omega_{rms}$) of the AM waveform for \textit{each} red/blue sideband when brought into resonance with the carrier transition is normalized by the Lamb-Dicke factor ($\eta \approx 0.08$) and the gate duration ($\tau_{gate} = 225\mu$s). Gate power is used as a proxy for crosstalk, and the ion$\rightarrow$qubit mapping is chosen to minimize the cost matrix.}
\label{fig:gate coupling}
\end{figure}

In \cref{fig:gate coupling}, we present the power requirements for the gates in our system. We first optimize the frequency of each gate across the mode spectrum to find pulse solutions that are robust to mode errors of $<1$~kHz. Once the frequency is fixed for each gate, we calculate the root-mean-square (RMS) Rabi frequency ($\Omega_{rms}$) of the AM waveform for \textit{each} red/blue sideband when brought into resonance with the carrier transition. In our system, we use equal Rabi frequencies to drive both ions $i,j$ in the gate ($\Omega_{i,rms}=\Omega_{j,rms}$), although this need not be case. The Lamb-Dicke factor ($\eta \approx 0.08$) converts carrier Rabi frequency to sideband frequency and this factor is normalized by the gate duration ($\tau_{gate} = 225\mu$ s). Using \cref{fig:gate coupling} as a cost matrix, we manually optimize the mapping so that the required gates in the circuit minimize the total cost. In general, we observe that each half of the chain has strong coupling to itself, and the two halves of the chain couple well to each other as long as symmetry of the chain is obeyed (e.g., gates where the ions are with both odd or both even integer offsets from the center of the chain couple well, but mixed even and odd integer offsets do not). We further note that when considering the full stabilizer circuit, it is preferable to use ion 8 as a data qubit (maximum 4 gates) than as an ancilla qubit (6 gates). With these considerations, we arrived at the ion$\rightarrow$qubit mapping displayed in Fig. 1 of the main text.

\new{
\subsection*{Magic State Fidelity}
To calculate the fidelity of our magic state preparation circuit for the state $\logical{H_x}$, we can compute the fidelity between a mixed state $\rho$, which represents the experimentally prepared state, and the ideal pure state $\ket{\psi}$ as
\begin{equation}
    \begin{split}
        F &= \langle \psi | \rho | \psi \rangle\\
        &= \mbox{Tr}[\rho |\psi\rangle\langle\psi |]\\
        &= \frac{1}{2}\left(1 + \langle X\rangle_\rho \langle X\rangle_\psi + \langle Y\rangle_\rho \langle Y\rangle_\psi + \langle Z\rangle_\rho \langle Z\rangle_\psi\right)\\
        &= \frac{1}{2}\left(1 + \langle X\rangle_\rho \frac{1}{\sqrt{2}} + \langle Z\rangle_\rho \frac{1}{\sqrt{2}}\right).
    \end{split}
\end{equation}

The expectation values $\langle Z\rangle_\rho$ and $\langle X\rangle_\rho$ can be extracted by measuring logical Z operator before and after a logical $Y_L(\pi/2)$ operation. This analysis leads to the following fidelities as shown in Table \ref{tab:Hx Fidelities}. 
\begin{table}[H]
    \centering
    \begin{tabular}{c|c}
         Processing Technique & Fidelity\\
         \hline
         Raw & $0.85\pm 0.01$\\
         Correction & $0.972\pm 0.012$\\
         Detection & $0.98 \pm 0.013$
    \end{tabular}
    \caption{Fidelities for the $\logical{H_x}$ magic state preparation circuit under different processing techniques.}
    \label{tab:Hx Fidelities}
\end{table}

However the same procedure cannot be applied to the $\logical{H_y}$ state, as the [[9,1,3]] Bacon-Shor code does not allow for fault-tolerant measurement in the logical Y basis. Using the constraint
$$\langle X\rangle^2 + \langle Y\rangle^2 + \langle Z\rangle^2 \leq 1$$
we can only numerically bound the fidelity of the $\logical{H_y}$ state to the range $0.75 \leq F \leq 0.99$. However, we argue that the fidelities for preparing $\logical{H_x}$ and $\logical{H_y}$ should be very similar, as the preparation circuit only differs in the phase of a single qubit gate, a quantity which we control to $\approx400\mu$rad limited by the AWG bit depth. This argument is further strengthened by the results of our single qubit benchmarking. Thus the $\logical{H_y}$ state fidelity should be very similar to values shown in Table \ref{tab:Hx Fidelities}.

}
\subsection*{Logical \Ttwostar fits}

The Ramsey fringe amplitudes shown in Fig.~2c are calculated by fitting a curve to a logical Ramsey experiment at each wait time. The data is taken by preparing a $\logical{+}$ state as shown in Eq.~1, waiting some amount of time $t$, applying varying $RZ(\theta)$ gates to every qubit and then measuring in the logical $X$ basis via a transversal $RY_L(-\pi/2)$ . Here we will explain the theoretical fits used for raw, corrected, and detected data processing techniques.

Firstly, as shown in Eq.~1, the logical $|+\rangle$ state we use is composed of three GHZ states $\frac{1}{\sqrt{2}}(|000\rangle + |111\rangle)$. Due to the structure of these states, if a $RZ(\theta)$ gate is applied to each qubit, the three gates will coherently combine, and the end result will be the same as if a $RZ(3\theta)$ gate had been applied on any single qubit. By considering this simplification we can reduce the number of error cases we must consider.

For the 'raw' processing case, any $Z$ error flips the logical output. As a result the cases where $1$ or $3$ errors occur lead to $\logical{-}$ states, while while cases with $0$ or $2$ errors lead to $\logical{+}$. Consequently the expectation value of $X_L$ can be thought of as the squared amplitude of cases which lead to $\logical{+}$, subtracted by the squared amplitude of cases which result in $\logical{-}$. This results in a curve
\begin{equation*}
        \langle X_L \rangle = \cos(3\theta/2)^6 - 3\cos(3\theta/2)^4\sin(3\theta/2)^2 + 3\cos(3\theta/2)^2\sin(3\theta/2)^4 - \sin(3\theta/2)^6 = \cos(3\theta)^3.
\end{equation*}

In the 'corrected' processing case, the state can tolerate a single error without having its logical information corrupted. As a result error cases with $0$ or $1$ errors lead to $\logical{+}$, while $2$ or $3$ lead to $\logical{-}$. This results in the curve
\begin{equation*}
        \langle X_L \rangle = \cos(3\theta/2)^6 + 3\cos(3\theta/2)^4\sin(3\theta/2)^2 - 3\cos(3\theta/2)^2\sin(3\theta/2)^4 - \sin(3\theta/2)^6.
\end{equation*}

Lastly the 'detected' processing method is slightly more complex, as postselection means we must renormalize the expectation value. The case with $0$ errors leads to $\logical{+}$, while the case with $3$ errors leads to $\logical{-}$. Cases with $1$ or $2$ errors must set off at least one stabilizer, and as a result those runs will be removed from the dataset. As a result the probabilities must be renormalized, leading to the curve
\begin{equation*}
    \langle X_L \rangle = \frac{\cos(3\theta/2)^6 - \sin(3\theta/2)^6}{\cos(3\theta/2)^6 + \sin(3\theta/2)^6}.
\end{equation*}

In an experiment there will also be imperfections in the states due to errors beyond \Ttwostar dephasing, which we model to be a simple depolarization of each GHZ state with strength $p$. This corresponds to taking a state $|+\rangle_{GHZ} \rightarrow (1-p)|+\rangle\langle +|_{GHZ} + \frac{p}{2} (|+\rangle\langle +|_{GHZ} + |-\rangle\langle -|_{GHZ})$, where the second term, equal to the maximally mixed state on the space spanned by $|000\rangle$ and $|111\rangle$, has an expectation value of 0.

In the raw case, any depolarization of the GHZ states will lead to the expectation value going to zero, and as a result the only non-zero expectation values come about when no depolarization occurs. As a result the overall fringe pattern is simply scaled by a factor of $(1-p)^3$:

$$\langle X_L \rangle = A(1-p)^3\cos(3\theta)^3.$$

In the correction case the stabilizers are able to identify and correct a single depolarization error. This leads to different cases for when depolarization occurs and when they do not, which when collated lead to:
\begin{equation*}
    \begin{split}
        \langle X_L \rangle = &\left[ (1-p)^3 + 3(1-p)^2 p + \frac{3p^2(1-p)}{2}\right] (\cos(3\theta/2)^6 - \sin(3\theta/2)^6)\\
        + &\left[ 3(1-p)^3 + 3(1-p)^2 p + \frac{3p^2(1-p)}{2}\right] (\sin(3\theta/2)^2\cos(3\theta/2)^4 - \sin(3\theta/2)^4\cos(3\theta/2)^2).
    \end{split}
\end{equation*}

The most complex case is the detection case. Individual depolarizations each contribute a $\frac{1}{2}$ chance of setting off a stabilizer, and when they do not the coherent rotations on the other qubits produce similar behaviors to the ideal detection case, but only on the non-depolarized qubits. This leads to the equation:
\begin{equation*}
    \begin{split}
        \langle X_L \rangle = A\bigg[ (1-p)^3&\left(\frac{\cos(3\theta/2)^6 - \sin(3\theta/2)^6}{\cos(3\theta/2)^6 + \sin(3\theta/2)^6}\right)\\
        + \frac{3p(1-p)^2}{2}&\left(\frac{\cos(3\theta/2)^4 - \sin(3\theta/2)^4}{\cos(3\theta/2)^4 + \sin(3\theta/2)^4}\right)\\
        + \frac{3p^2(1-p)}{4}&\left.\left(\frac{\cos(3\theta/2)^2 - \sin(3\theta/2)^2}{\cos(3\theta/2)^2 + \sin(3\theta/2)^2}\right) \right].
    \end{split}
\end{equation*}

\begin{figure}[H]
\centering
\includegraphics[width=1\textwidth]{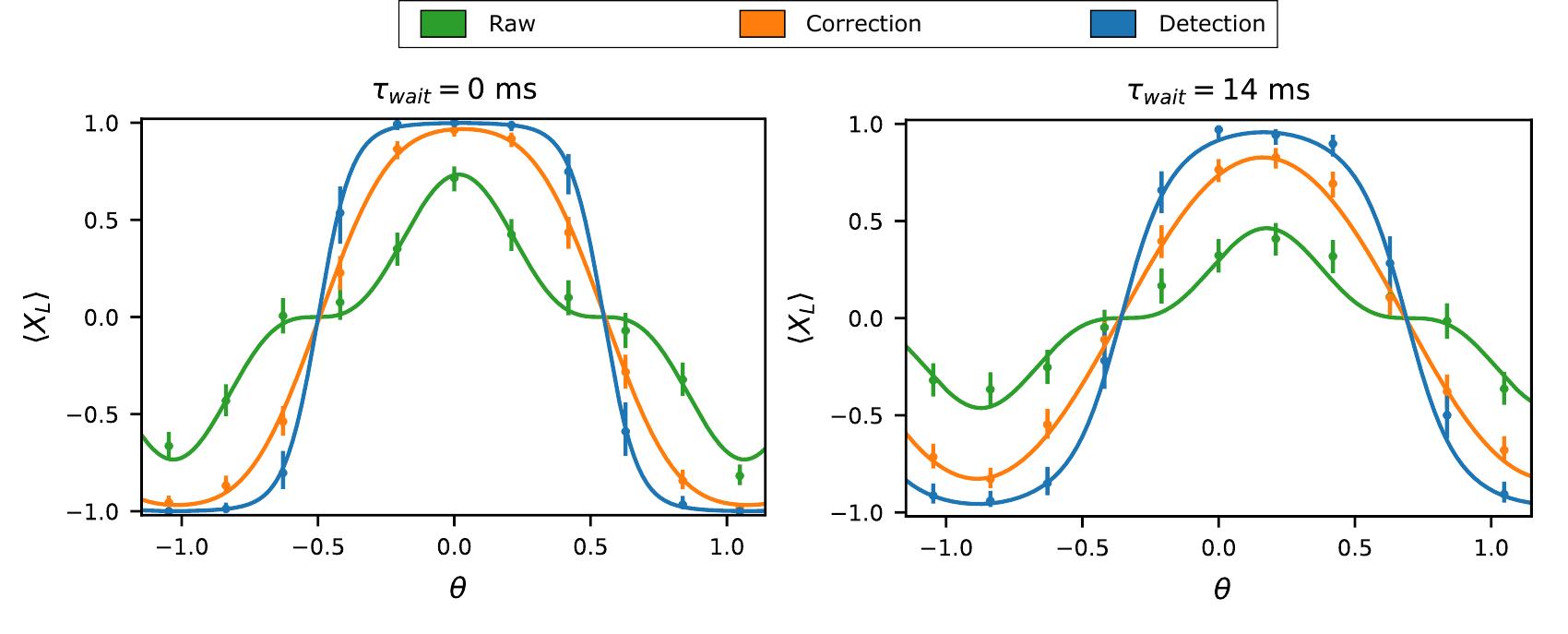}
\caption{\textbf{Examples of \Ttwostar fits.} These plots exemplify logical Ramsey fringe fitting at two different wait times, 0ms (left) and 14ms (right). At short wait times (left), the data shows characteristics of error correction, such as the flattened top for error-detection. At longer wait times (right), the amplitude of each curve decreases due to \Ttwostar, but also the flat features of the curves blur due to GHZ depolarization. In both cases, the error model well matches the experimental data. Error bars are the $95\%$ binomial proportion confidence interval}
\label{fig:t2star fits}
\end{figure}

These models well describe the experimental data, as shown in \cref{fig:t2star fits}.

\subsection*{Physical \Ttwostar}
\begin{figure}[H]
\centering
\includegraphics[width=0.5\textwidth]{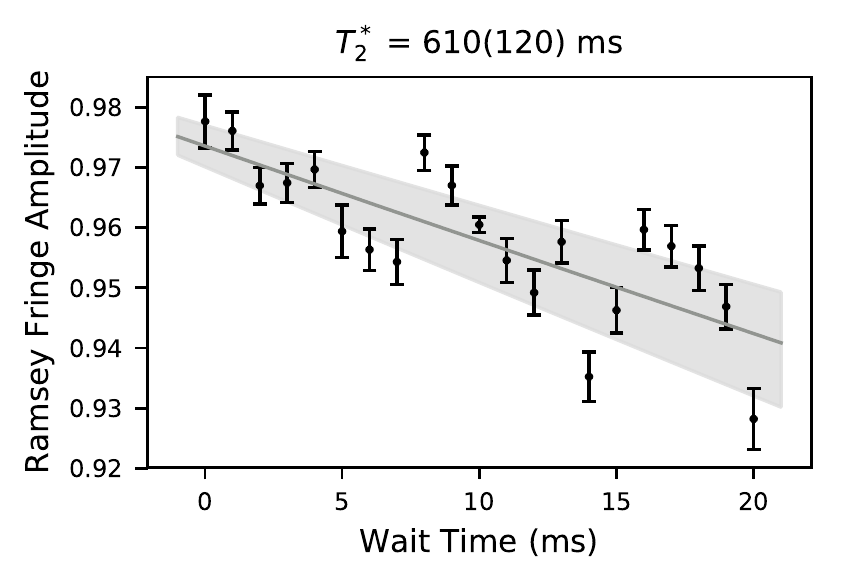}
\caption{\textbf{Raman \Ttwostar for a physical qubit in a 15-ion chain.} Ramsey fringes with variable wait times are fit to a sinusoud to extract the contrast. For each data point, the error bars represent the 95\% confidence interval from a maximal likelihood estimation fit of a sinuosoid. The contrast is fit to a decaying exponential $Ae^{-\tau_{\rm{wait}}/\Ttwostar}$, with fit value $\Ttwostar = 610(120)$~ms. The shaded region indicates the $1\sigma$ uncertainty in the exponential least-squares fit. }
\label{fig:phys t2star}
\end{figure}

To understand the phase flip errors in the logical qubit, we measure the \Ttwostar of the physical qubits in a chain of 15 ions. This is accomplished via a laser Raman Ramsey sequence on the center ion, $RY(\pi/2)-\tau_{\rm{wait}}-RZ(\theta)-RY(-\pi/2)$, with no echoes. At each wait time $\tau_{\rm{wait}}$, the phase $\theta$ is swept, and the resulting data is fit to a sinusoid to extract the contrast. The Ramsey contrast is fit to a decaying exponential $Ae^{-\tau_{\rm{wait}}/\Ttwostar}$ to extract \Ttwostar. The results of this experiment are shown in \cref{fig:phys t2star}. We find $\Ttwostar = 610(120)$~ms for a physical qubit in a chain of 15 ions.  We attribute the physical qubit decoherence primarily to control noise, rather than to fundamental qubit decoherence. In particular, we note that there are features of revivals at $\approx 8$ ms and $16$ ms, corresponding to noise at $\approx 125$~Hz. We assign this to mechanical fluctuations (e.g., fans) that shift the standing wave of the optical Raman beams relative to the ions. This effect can be mitigated by switching to a ''phase-insensitive'' configuration \cite{inlek2014quantum}.

\begin{figure}[H]
\centering
\includegraphics[width=0.5\textwidth]{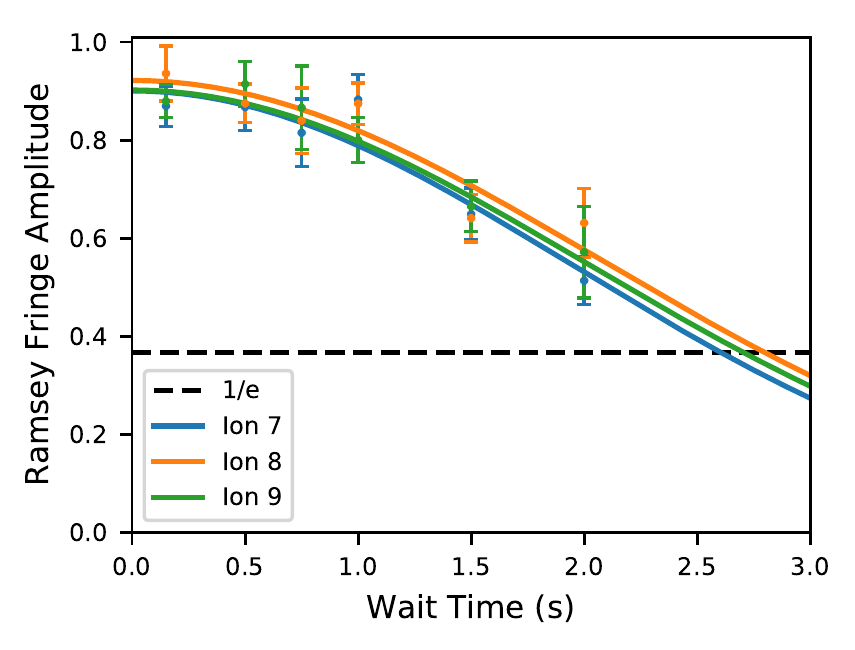}
\caption{\textbf{Microwave $T_2$ for a physical qubit in a 15-ion chain.}  An $(XY)^N$ dynamical decoupling sequence is used to suppress magnetic field noise. Ramsey fringes with variable wait times are fit to a sinusoud to extract the contrast. For each data point, the error bars represent the $1\sigma$ confidence interval from a least-squares fit of a sinuosoid. The contrast is fit to a Gaussian decay $Ae^{-(\tau_{\rm{wait}}/T_2)^2}$, with average fit value $T_2 = 2.84(16)$~s. }
\label{fig:MW t2star}
\end{figure}

To investigate the degree to which control errors impact our physical \Ttwostar qubit decoherence, we perform microwave Ramsey experiments, which are not sensitive to optical beam path fluctuations. Additionally, we suppress magnetic field inhomogeneity using a dynamical decoupling technique that applies $\pi$-pulses with alternating $90\degree$ phase offsets, commonly known as an $(XY)^N$ pulse sequence \cite{gullion1990new}, to periodically refocus the qubit spin. Due to a strong microwave drive field gradient along the ion chain, the $\pi$-pulse times are only well calibrated for three neighboring ions in the chain, which we center on the middle ion (8) in the chain. At each wait time, the phase of the fringe is swept, and the resulting data is fit to a sinusoid to extract the contrast. The resulting data is shown in \cref{fig:MW t2star}. We observe that the resulting decay is better fit to a Gaussian ($Ae^{-(\tau_{\rm{wait}}/T_2)^2}$), compared to an exponential decay, with average $T_2 = 2.84(16)$~s. The coherence time of this echo experiment is limited by residual magnetic field noise, which can be improved by operating our qubit in a lower bias-field or by using magnetic shielding. We note that $T_2 > 1$ hour has been achieved in \Yb \cite{wang2020single}.

\begin{figure}[H]
\centering
\includegraphics[width=0.5\textwidth]{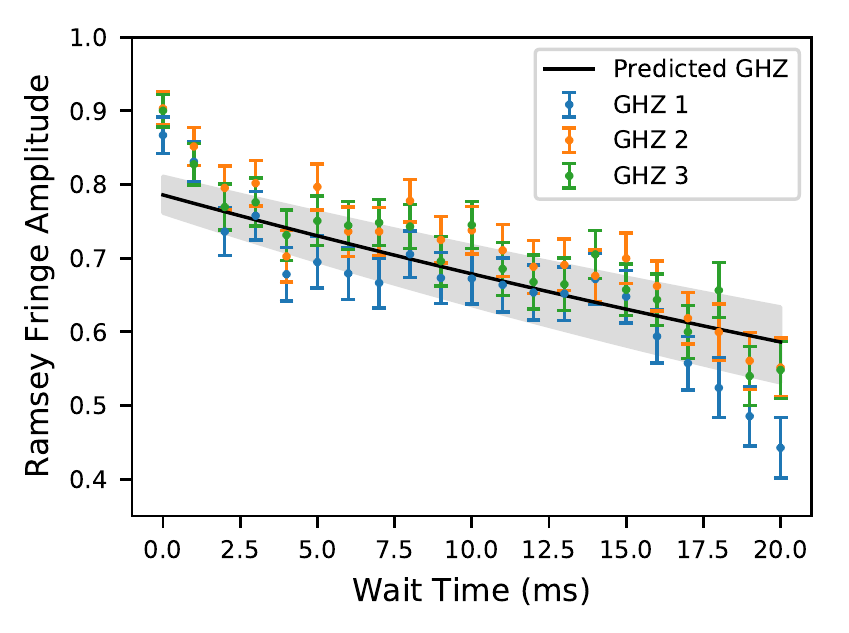}
\caption{\textbf{\Ttwostar for GHZ states.} Ramsey fringes for each independent GHZ state are analyzed from the logical \Ttwostar experiment in Fig. 2c of the main text. For each data point, the error bars represent the 95\% confidence interval from a maximal likelihood estimation fit of a sinuosoid. For the theoretical prediction, we apply the fitted dephasing noise in Fig. \ref{fig:phys t2star} to a numerical simulation of a three-qubit GHZ state. The shaded region indicates the $1\sigma$ uncertainty in the exponential least-squares fit in Fig. \ref{fig:phys t2star}, propagated through the numerical simulation.}
\label{fig:ghz t2star}
\end{figure}

The GHZ states that make up the Bacon-Shor code ($\ket{000}+\ket{111})$ are three times as sensitive to phase noise as our physical qubit. To understand the implication of the Raman \Ttwostar on the logical \Ttwostar, we run numerical simulations to extrapolate the measured phase noise to a GHZ state. We assume that the Pauli-$Z$ noise in the middle of the Ramsey sequence is Gaussian distributed with some width $\Delta_Z$. Using the fit from \cref{fig:phys t2star}, we can numerically solve for the width of the noise spectrum $\Delta_Z$. Once this value is found, we re-run the simulation with that noise spectrum on a three-qubit GHZ state to extract the predicted contrast. In \cref{fig:ghz t2star}, we compare this predicted value to the three individual GHZ states measured in the logical qubit experiment. We conclude that almost all the dephasing in the logical qubit that we observe is explained by the observed \Ttwostar- decay in the physical qubit. We note that this is the same experimental data presented in Fig. 2c of the main text, just post-processed to analyze individual GHZ states rather than to perform error-correction.

\subsection*{Bacon-Shor gauge operators}
The Bacon-Shor code is an example of a subsystem quantum error correcting code. These codes have additional quantum degrees of freedom which are not protected to the same distance as the logical degree of freedom. In the [[9,1,3]] Bacon-Shor code, there are 4 additional degrees of freedom referred to as \textit{gauge qubits}. One basis for the gauge qubits corresponds to fixing 4 constraints on the eigenvalues of the operators shown in the table below.
\begin{center}
\begin{tabular}{ c|c } 
$X$-gauges & $Z$-gauges\\
 \hline
 $X_1X_2$ & $Z_1Z_4$\\
 $X_4X_5$ & $Z_2Z_5$\\
 $X_7X_8$ & $Z_3Z_6$\\
 $X_2X_3$ & $Z_4Z_7$\\
 $X_5X_6$ & $Z_5Z_8$\\
 $X_8X_9$ & $Z_6Z_9$\\
\end{tabular}
\end{center}
It should be noted that this is not an independent set of operators because the stabilizers of the code, which are products of gauges, already have their eigenvalues fixed to $+1$. As such, if the $X$-gauges $X_1X_2$ and $X_4X_5$ both have eigenvalue $+1$ on a given logical state, then the eigenvalue of $X_7X_8$ will also be $+1$. We refer to a state in which all $X(Z)$-type gauge operators have eigenvalue $+1$ as the $\ket{\overline{X}}_G (\ket{\overline{Z}}_G)$ gauge. It should be noted that these gauge operators do not commute, so these two gauges are mutually exclusive. When decoding the Bacon-Shor code, we can only identify operators up to a product of gauges. When the correction is applied, we may have inadvertently applied a gauge operator to the logical state. This leaves our logical qubit unaffected, but will alter the state of the gauge qubits.

Logical Pauli operators on a subsystem code decompose as a tensor product of operations on the logical and gauge degrees of freedom. When a logical Pauli operator that acts non-trivially on the gauge subsystem is used to generate a continuous unitary operator, it will entangle the logical and gauge subsystems. As these gauge subsystems are less protected than the logical subsystem, the information will be less protected. Consequently, one must design continuous logical operators around logical Pauli operations that commute with the entire gauge group, ensuring that it acts trivially on the gauge subsystem.

\subsection*{Fit values for logical gate operations}
In Table.~\ref{tab:rabifits A} and \ref{tab:rabifits G}, we report the numerical values obtained from 
fitting the logical gate operations as displayed in Fig. 4d of the main text. In addition to the FT transversal gate and the nFT continuous gate, we also provide the fit values for a similar sampling of states in the logical $XZ$-plane generated from the nFT encoding circuit (Fig. 2a of main text). Error values are reported as the $1\sigma$ from a Gaussian approximation to a maximal likelihood estimation fit. The Gaussian approximation fails when the fit parameters are at or equal to their extrema (e.g., when $A\simeq 1$), in which case asymmetric error bars are given by the notation $(^{1\sigma \rm{upper}}_{1\sigma \rm{lower}})$.

\begin{table}[H]
\centering
\begin{tabular}{c c c c}
    \textbf{Amplitude ($A$)} &  Raw & Error Correction & Error Detection\\
    \hline
    FT Gate & $0.923(6)$ & $0.996(1)$ & $1.000 (^{0}_{3})$  \\
    nFT Gate & $0.89(1)$ & $0.87(1)$ & $1.00(^{0}_{1})$ \\
    nFT Encoding & $0.78(1)$ & $0.982(5)$ & $0.997(2)$ \\

\end{tabular}
\caption{Numerical values for the fit parameter $A$.}
\label{tab:rabifits A}
\end{table}

\begin{table}[H]
\centering
\begin{tabular}{c c c c}
    \textbf{Gate Error ($\Gamma$)} &  Raw & Error Correction & Error Detection\\
    \hline
    FT Gate & $0.044(3)$ & $0.0027(7)$ & $0.0002(1)$  \\
    nFT Gate & $0.015(5)$ & $0.000(4)$ & $0.011(2)$ \\
    nFT Encoding & $0.001(6)$ & $0.000 (2)$ & $0.001(1)$ \\

\end{tabular}
\caption{Numerical values for the fit parameter $\Gamma$.}
\label{tab:rabifits G}
\end{table}

\new{
\subsection*{Stabilizers on different input states}
In the main text (Fig 3a), we describe an experiment to measure $X$-type errors on the logical qubit state propagated from a $Z$-type errors on an ancilla during a FT $X$ stabilizer measurement circuit. In that experiment, there was no detectable change in the error rate. However, because the input and measured state was $\logical{0}$, the measurement is only sensitive to $X$-type errors. To check if the stabilizer is introducing $Z$-type errors into the logical qubit, we perform the same experiment (with no artificial error added) on the $\logical{+}$ input state. $\logical{+}$ is measured in the $\ev{X}_L$ basis by a transversal $Y_L(-\pi/2)$ gate that maps $\ev{X}_L \rightarrow \ev{Z}_L$. We also check a $Z$ stabilizers on both input states. The results are shown in Table \ref{tab:stab_bases}. We note that from the logical \Ttwostar experiment, we expect the error on the $\logical{+}$ state to increase by $1.8\%$ over the $\approx1.5$ms required to measure the stabilizer, which is not included in the baseline encoding error below. 

\begin{table}[H]
\centering
\begin{tabular}{c c c c}
    Input State & Baseline Encoding & FT $Z$ Stabilizer & FT $X$ Stabilizer\\
    & (\% Error)& (\% Error)& (\% Error)\\
    \hline
    $\logical{0}$ & $0.23(13)$ & $0.41(10)$ & $0.20(13)$  \\
    $\logical{+}$  & $0.45(11)$ & $3.3(3)$ & $2.1(2)$ \\
\end{tabular}
\caption{Logical error rates for both $Z$ and $X$ FT stabilizers on different logical input states, after error-correction. Uncertainties are $1\sigma$ from the binomial distribution.}
\label{tab:stab_bases}
\end{table}
}

\subsection*{Extended stabilizer results}
\begin{figure}[H]
\centering
\includegraphics[width=\textwidth]{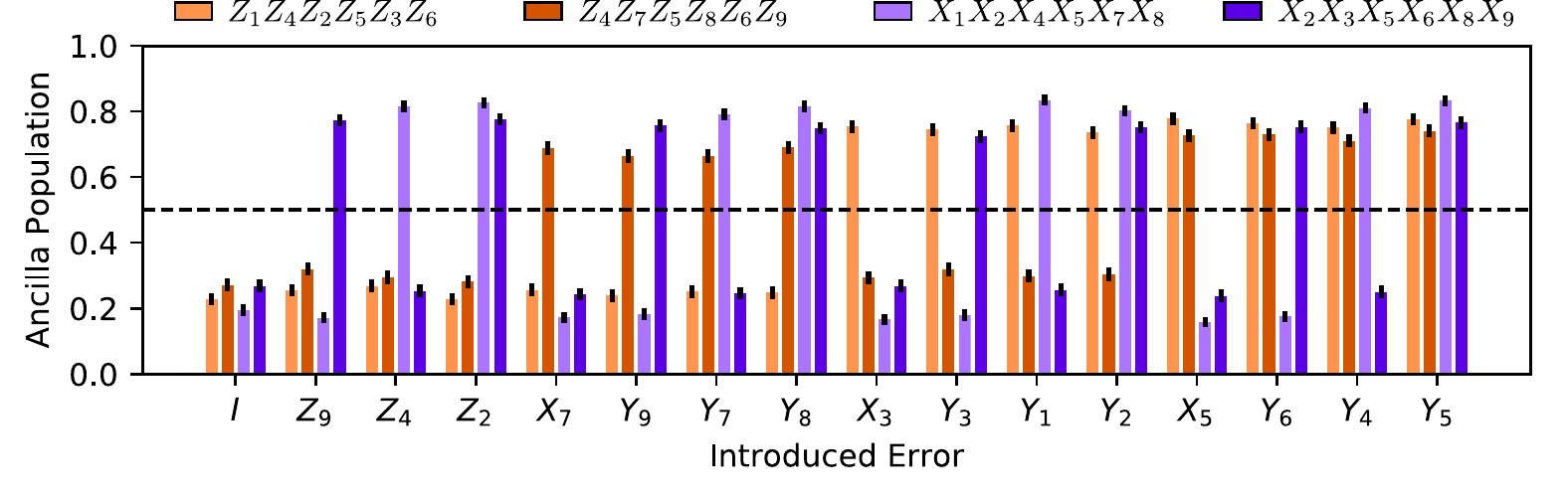}
\caption{\textbf{Full stabilizer measurement results.} A complete set of errors are introduced to create all possible output bit strings of the ancilla qubits. Error bars are the $95\%$ binomial proportion confidence interval.}
\label{fig:full stab}
\end{figure}

In Fig. 3 of the main text, we presented a representative sample of artificially introduced errors and the corresponding ancilla qubit populations. Here in \cref{fig:full stab}, we present a full set of errors that produce all of the possible ancilla qubit output bit strings.

For a given error, each stabilizer measurement yields a deterministic eigenvalue measurement (e.g., $\{+1,-1,-1,+1)\}$ that is mapped to the ancilla qubit state (e.g., $\{0,1,1,0\}$). Defining the error as the difference between the expected ancilla bit string and the measured populations, and averaging across all the artificial errors, we obtain the following total error for each stabilizer measurement: 

\begin{table}[H]
\centering
\begin{tabular}{c c}
    \textbf{Stabilizer} & \textbf{Total Error ($\epsilon_{S_i}$)} \\
    \hline
    $S_1 = Z_1 Z_4 Z_2 Z_5 Z_3 Z_6$ & $0.244(3)$ \\
    $S_2 = Z_4 Z_7 Z_5 Z_8 Z_6 Z_9$ & $0.298(6)$ \\
    $S_3 = X_1 X_2 X_4 X_5 X_7 X_8$ & $0.179(3)$ \\
    $S_4 = X_2 X_3 X_5 X_6 X_8 X_9$ & $0.248(3)$ \\
\end{tabular}
\end{table}

In this experiment, the stabilizers are measured in the order $S_3, S_4, S_1, S_2$. We note from the data presented in Fig. 2 of the main text, the raw encoding of the $\logical{0}$ state has a base $\epsilon_{enc} = 0.038(2)$ error, which we assume is isotropic in the sense that all stabilizer measurements should see the error equally. Additionally, stabilizer measurements will detect errors introduced by itself or previous stabilizer measurements, which we assume to be isotropic as well. The per stabilizer error can be calculated by the differential error between successive stabilizer measurements. We calculate $\epsilon_Z = 0.064(7)$ per $Z$-stabilizer (avg. $98.9\%$ gate fidelity) and $\epsilon_X = 0.069(5)$ per $X$-stabilizer (avg. $98.8\%$ gate fidelity). Finally, we observe an error offset on the $X$-stabilizers relative to the $Z$-stabilizers of $\epsilon_{\Ttwostar} = 0.072(5)$, consistent with a $Z$-type error caused by the logical qubit dephasing (\Ttwostar) over the wall-clock time it takes to measure the $X$-stabilizers ($\approx 3$~ms), as presented in Fig. 2c of the main text. In conclusion, we find that the total stabilizer measurement error for each ancilla qubit is well explained by the following error model:

\begin{equation}
\begin{aligned}
    \epsilon_{S_1} &= \epsilon_{enc} + 2\epsilon_{X} + \epsilon_{Z}\\
    \epsilon_{S_2} &= \epsilon_{enc} + 2\epsilon_{X} + 2\epsilon_{Z}\\  
    \epsilon_{S_3} &= \epsilon_{enc} + \epsilon_{\Ttwostar} + \epsilon_{X} \\ 
    \epsilon_{S_4} &= \epsilon_{enc} + \epsilon_{\Ttwostar} + 2\epsilon_{X}\\
\end{aligned}
\end{equation}

\new{
\subsection*{Native Gate Circuit Library}

\begin{figure}[H]
\centering
\includegraphics[width=0.5\textwidth]{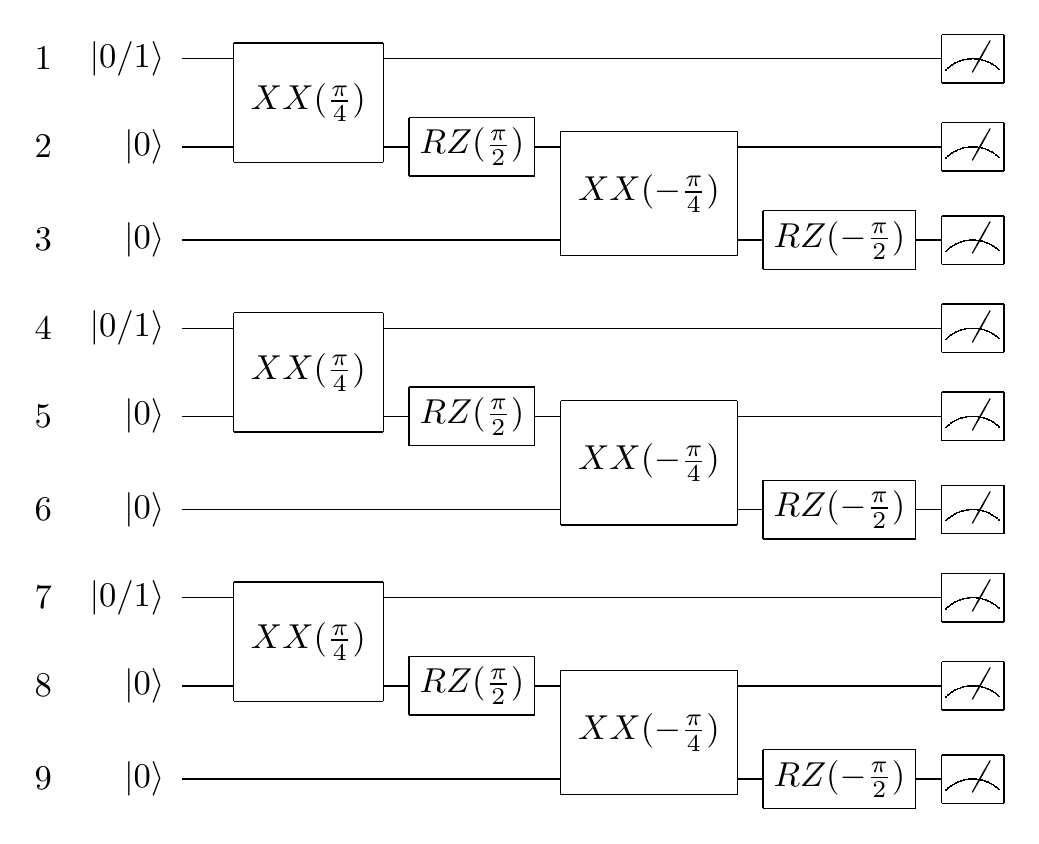}
\caption{\textbf{Native gate compilation for encoding $\logical{0/1}$.} To create the $\logical{0/1}$ state, qubits 1, 4, and 7 should be initialized in the $\ket{0/1}$ state, respectively. Corresponds to the blue Clifford circuit in Fig.~2a and the blue data in Fig.~2b of the main text/}
\label{fig:logical-01}
\end{figure}

\begin{figure}[H]
\centering
\includegraphics[width=0.65\textwidth]{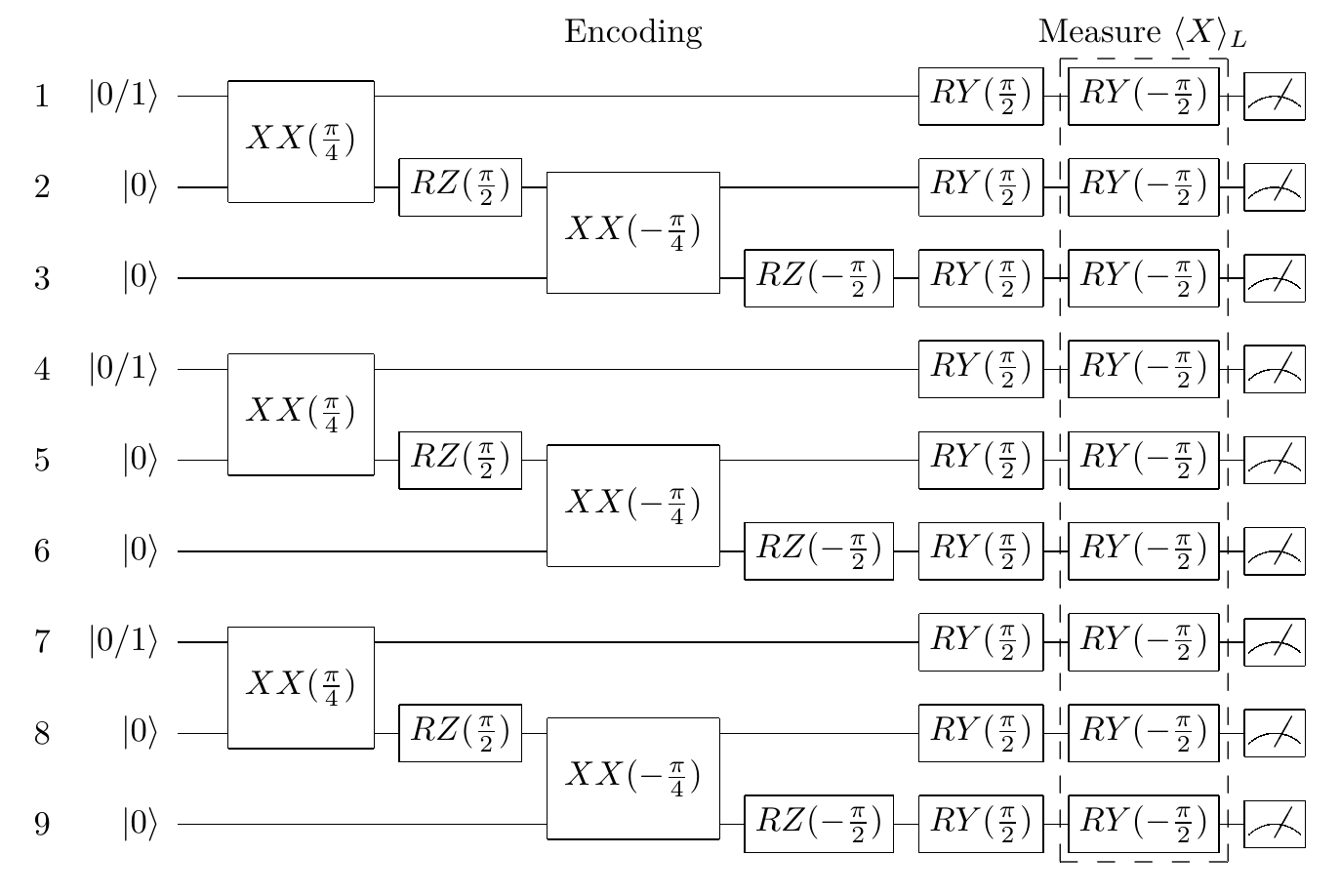}
\caption{\textbf{Native gate compilation for encoding $\logical{+/-}$.} To create the $\logical{+/-}$ state, qubits 1, 4, and 7 should be initialized in the $\ket{0/1}$ state, respectively. We note that we do not compile out the successive $RY(\pm \frac{\pi}{2})$ gates on each qubit, as this would create a circuit identical to \ref{fig:logical-01} - this explains the additional error on $\logical{+/-}$ states relative to $\logical{0/1}$. Corresponds to the blue circuit in Fig.~2a and the blue data in Fig.~2b of the main text.}
\label{fig:logical-pm}
\end{figure}

\begin{figure}[H]
\centering
\includegraphics[width=0.8\textwidth]{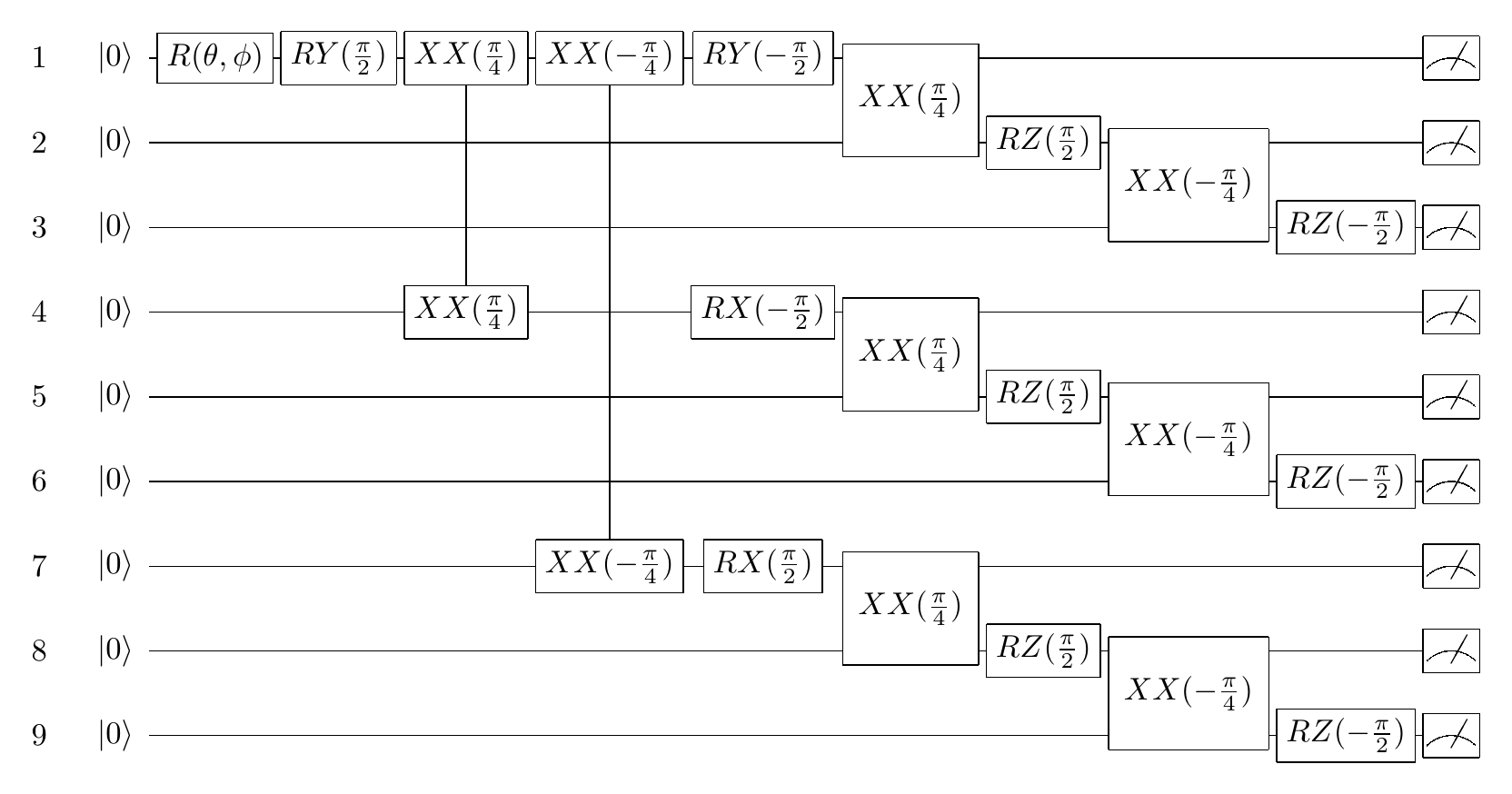}
\caption{\textbf{Native gate compilation for nFT encoding of arbitrary logical states.} This circuit can create an arbitrary $\logical{\psi}$ state controlled by the initial $R(\theta,\phi)$ gate on qubit 1, with $\ket{\psi}_1 = R(\theta,\phi)\ket{0}_1$. Corresponds to the red circuit in Fig.~2a and the red data in Fig.~2b of the main text. This circuit is also used for initial encoding of the magic states in Fig.~c-d of the main text.}
\label{fig:directprep}
\end{figure}

\begin{figure}[H]
\centering
\includegraphics[width=0.9\textwidth]{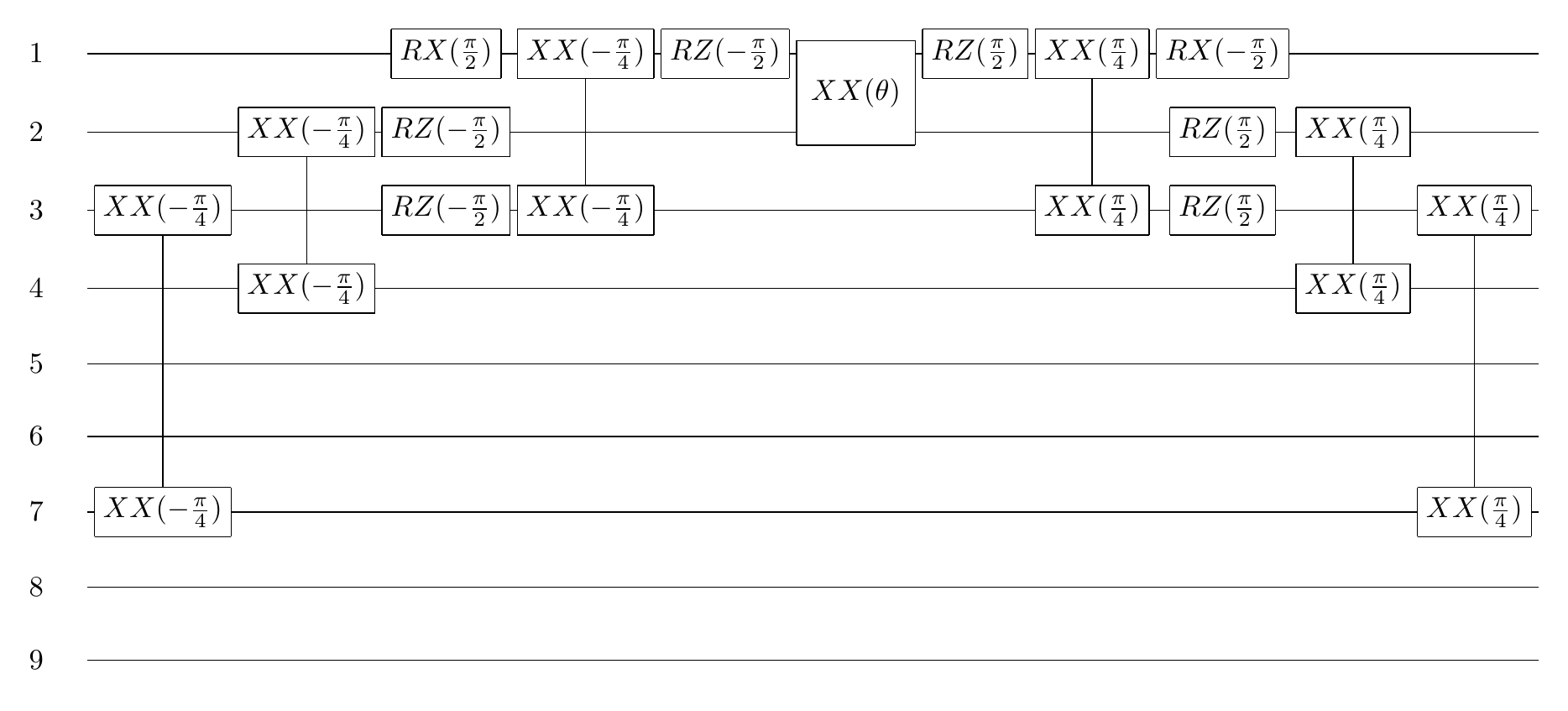}
\caption{\textbf{Native gate compilation for nFT continuous rotation.} This circuit implements a $Y_L(\theta) = Y_1Z_2Z_3X_4X_7(\theta)$ rotation controlled by an $XX(\theta)$ gate on qubits 1 and 2. Corresponds to the red circuit in Fig.~3b and the red data in Fig.~3d-e of the main text.}
\label{fig:continuous}
\end{figure}

\begin{figure}[H]
\centering
\includegraphics[width=\textwidth]{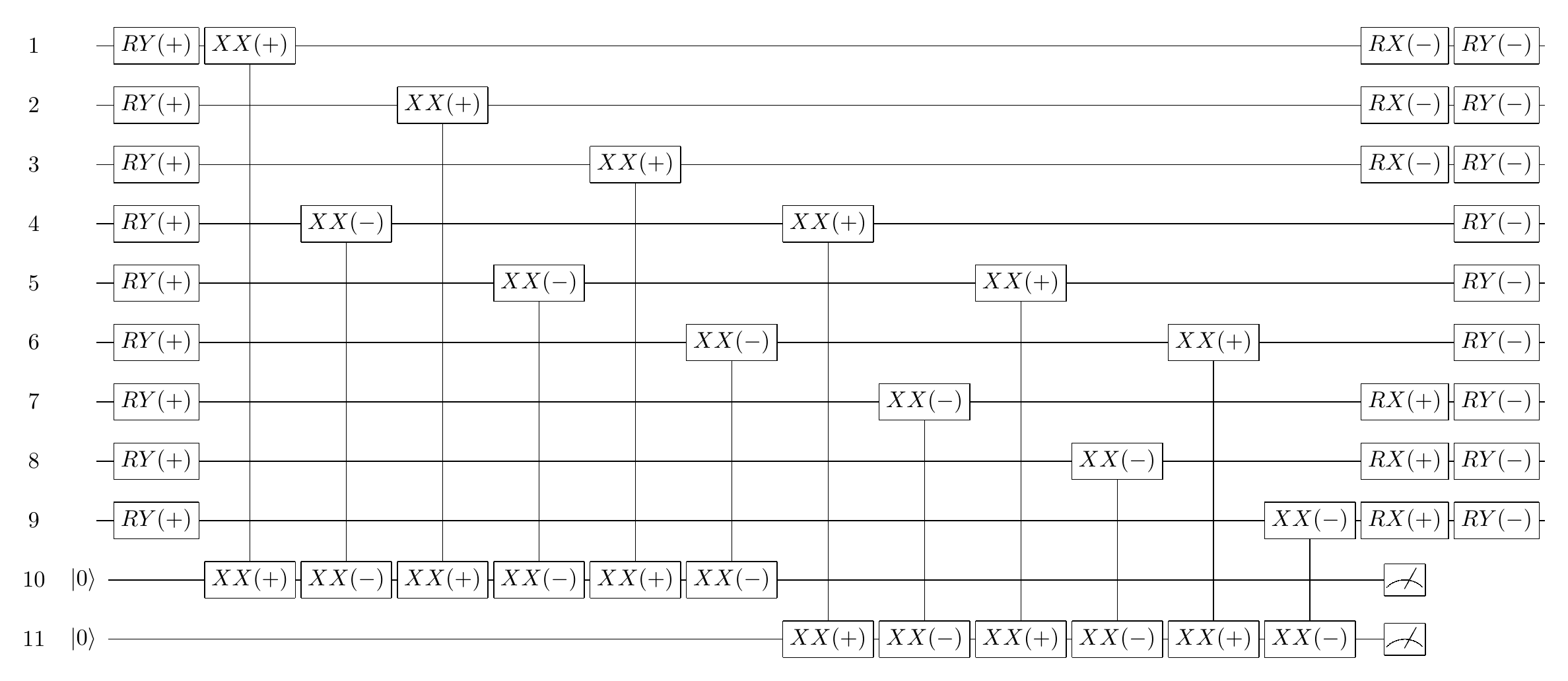}
\caption{\textbf{Native gate compilation for Z stabilizer measurement.} Here, the angle $(+/-)$ is shorthand for $(\pm\frac{\pi}{2})$ or $(\pm\frac{\pi}{4})$ for single- or two-qubit gates, respectively. When combined with encoding (Fig.~\ref{fig:logical-01}) and the X stabilizer (Fig.~\ref{fig:Xstab}), the combined circuits measure the full syndrome (Extended Data Fig.~1a). }
\label{fig:Zstab}
\end{figure}

\begin{figure}[H]
\centering
\includegraphics[width=\textwidth]{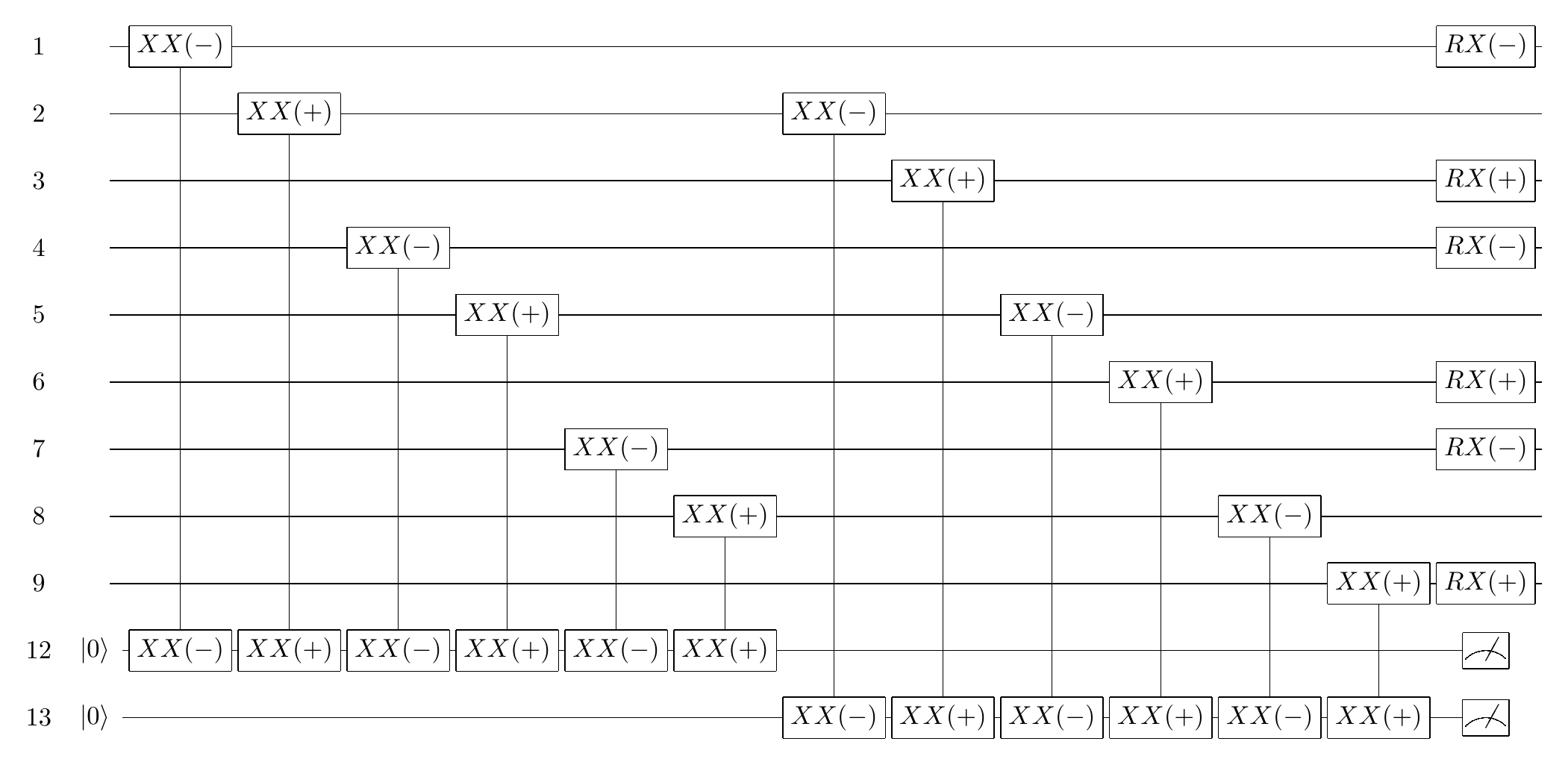}
\caption{\textbf{Native gate compilation for FT X stabilizer measurement.} Here, the angle $(+/-)$ is shorthand for $(\pm\frac{\pi}{2})$ or $(\pm\frac{\pi}{4})$ for single- or two-qubit gates, respectively. Corresponds to the circuit in Extended Data Fig.~1c and for data presented in Fig.~4a of the main text.}
\label{fig:Xstab}
\end{figure}
}

{\bibliography{bibliography}}

\end{document}